\documentclass[journal]{IEEEtran}
\usepackage[utf8]{inputenc}
\usepackage{amsmath,amssymb,amsfonts}
\usepackage{algorithm} 
\usepackage{epstopdf}
\usepackage{subfigure}
\usepackage{textcomp}
\usepackage{bm}
\usepackage{cite}

\usepackage{amsmath,amsfonts}
\usepackage{amsmath}
\usepackage{algorithmic}
\usepackage{array}
\usepackage[caption=false,font=normalsize,labelfont=sf,textfont=sf]{subfig}
\usepackage{textcomp}
\usepackage{stfloats}
\usepackage{url}
\usepackage{verbatim}
\usepackage{graphicx}
\usepackage{booktabs}

\def\BibTeX{{\rm B\kern-.05em{\sc i\kern-.025em b}\kern-.08em
		T\kern-.1667em\lower.7ex\hbox{E}\kern-.125emX}}
\usepackage{balance}
\begin{document}
	
	\title{\huge Service Provisioning and Path Planning with Obstacle Avoidance for Low-Altitude Wireless Networks}
	\author{ Senning Wan, Bin Li, Hongbin Chen, and Lei Liu
		\thanks{ Senning Wan and Bin Li are with the School of Computer Science, Nanjing University of Information Science and Technology, Nanjing 210044, China (e-mail: 202412492192@nuist.edu.cn; bin.li@nuist.edu.cn).} 
		\thanks{Hongbin Chen is with the School of Information and Communication, Guilin University of Electronic Technology, Guilin 541004, China (e-mail: chbscut@guet.edu.cn).}
		\thanks{Lei Liu is with the Guangzhou Institute of Technology, Xidian University, Guangzhou 510555, China (e-mail: tianjiaoliulei@163.com).}
	}
	
	\setlength{\parskip}{0pt} 
	\maketitle
	\begin{abstract}
		This paper investigates the three-dimensional (3D) deployment of uncrewed aerial vehicles (UAVs) as aerial base stations in heterogeneous communication networks under constraints imposed by diverse ground obstacles. Given the diverse data demands of user devices (UDs), a user satisfaction model is developed to provide personalized services. In particular, when a UD is located within a ground obstacle, the UAV must approach the obstacle boundary to ensure reliable service quality. Considering constraints such as UAV failures due to battery depletion, heterogeneous UDs, and obstacles, we aim to maximize overall user satisfaction by jointly optimizing the 3D trajectories of UAVs, transmit beamforming vectors, and binary association indicators between UAVs and UDs. To address the complexity and dynamics of the problem, a block coordinate descent method is adopted to decompose it into two subproblems. The beamforming subproblem is efficiently addressed via a bisection-based water-filling algorithm. For the trajectory and association subproblem, we design a deep reinforcement learning algorithm based on proximal policy optimization to learn an adaptive control policy. Simulation results demonstrate that the proposed scheme outperforms baseline schemes in terms of convergence speed and overall system performance. Moreover, it achieves efficient association and accurate obstacle avoidance.
	\end{abstract}
	
	\section{Introduction}
    The upcoming sixth-generation (6G) wireless networks aim to provide massive connectivity and ultra-high-speed transmission for millions of interconnected devices~\cite{art1}. However, the deployment of terrestrial infrastructure in sparsely populated areas, such as deserts, mountains, and forests, faces significant challenges due to environmental and economic constraints.  Furthermore, terrestrial base stations (BSs) are often vulnerable to unexpected events caused by environmental factors, such as earthquakes, floods, and tsunamis. Therefore, establishing a reliable emergency communication system via temporary aerial networks has become a key objective in the evolution toward 6G~\cite{art2}. Uncrewed aerial vehicles (UAVs), due to their high mobility and rapid deployment capabilities, have been extensively adopted. When functioning as aerial BSs, UAVs can dynamically adjust their trajectories according to actual environmental conditions and terrain features, thereby enhancing system robustness and communication efficiency~\cite{art3,art4,art5}. 

   In practical scenarios, UAV flight is frequently constrained by other UAVs and ground obstacles, which not only increases the complexity of collision avoidance but also imposes critical challenges on communication quality~\cite{art6,art7}. In this work, we consider the three-dimensional (3D) trajectory optimization of UAVs. Specifically, for ground obstacles, UAVs are allowed to either bypass them laterally or fly over them to ensure safe flight. Moreover, considering that user devices (UDs) have heterogeneous requirements for data rates, we develop a user satisfaction model to characterize the performance of communication services. Given the aforementioned considerations, the UAV flight trajectories and communication decisions must be carefully designed to maximize user satisfaction.
  
   \subsection{Previous Work}
   According to Shannon's capacity formula, proper trajectory planning can significantly reduce the communication distance between UDs and UAVs, thereby enabling the fulfillment of high data rate requirements~\cite{tr2,tr3}. For example, Lu \textit{et al.}~\cite{tr2} studied the joint optimization of trajectory planning and communication design for multiple UAV-BSs, aiming to maximize the fairness of throughput among the ground users (GUs) requiring service. Shang \textit{et al.}~\cite{tr3} employed a double deep Q-network (DDQN) to derive reward parameters based on fairness and throughput, which interact with the dynamic channel environment to continuously update UAV positions and maximize the accumulated reward. In addition, proper resource allocation, such as UAV transmit power control, significantly affects system performance and user service quality~\cite{beamforming1,beamforming2}. For instance, Li \textit{et al.}~\cite{beamforming1} investigated a secure communication system assisted by reconfigurable intelligent surfaces (RIS), where the UAV trajectory, RIS passive beamforming, and transmit power of legitimate transmitters were jointly optimized to maximize the average worst-case secrecy rate. In~\cite{beamforming2}, Liu \textit{et al.} proposed a joint design of transmit beamforming and UAV trajectory for enhanced communications and radar sensing tasks.

   During UAV flight, although trajectory design considers practical factors such as UAV speed, initial/final positions, and energy efficiency, collision avoidance is crucial for ensuring UAV survivability and operational safety in low-altitude environments with mixed obstacles. Therefore, careful trajectory planning is necessary~\cite{avoid1,avoid2,avoid3,avoid4,avoid5,avoid6}. Singla \textit{et al.}~\cite{avoid1} utilized onboard image collection devices and neural networks to extract visual features and estimate distances between UAVs and obstacles, enabling effective obstacle avoidance. Zheng \textit{et al.}~\cite{avoid2} investigated the 3D UAV trajectory planning problem in data collection systems, aiming to collect data in minimal time while avoiding spatial obstacles. Similarly, Gao \textit{et al.}~\cite{avoid3} considered ground terminal mobility, 3D building avoidance, and the practical assumption that UAVs can fly between buildings to minimize the total time cost. To achieve this objective, a multi-step dueling DDQN method was adopted, allowing the agent to interact with the environment and iteratively refine its motion policy. However, the aforementioned works mainly focus on single-UAV scenarios, while obstacle avoidance in multi-UAV environments remains underexplored. Gao \textit{et al.}~\cite{avoid4} proposed an energy-efficient velocity control algorithm for massive UAV networks based on mean field game theory, jointly addressing energy consumption, channel capacity, and obstacle avoidance. In~\cite{avoid5}, Wang \textit{et al.} investigated efficient and accurate solutions for path-following, obstacle sensing, and avoidance subtasks, as well as their conflict-free fusion and scheduling. Additionally, Gao \textit{et al.}~\cite{avoid6} proposed a potential game-based multi-agent deep deterministic policy gradient (DDPG) method to optimize obstacle-avoidance trajectories for multiple UAVs. In summary, research on 3D obstacle avoidance for multi-UAV scenarios remains limited and warrants further investigation. 
\begin{table*}[t]
	\centering
	\caption{Differences in Our Scheme Compared to the Most Relevant Schemes}
	\resizebox{\textwidth}{!}{
		\renewcommand{\arraystretch}{0.66}
		\begin{tabular}{@{}ccccccc@{}} 
			\toprule
			Schemes & Heterogeneous User & Multiple obstacles & Energy limitation & 3D trajectory & User satisfaction & Learning Methodology  \\
			\midrule
			{\cite{tr3}}   & -- & -- & \checkmark &  -- & -- & DRL \\
			{\cite{avoid2}}   & -- & -- & \checkmark & -- & -- & BCD  \\
			{\cite{avoid3}}   & -- & -- & -- &  \checkmark & -- & DRL \\
			{\cite{statis1}}  & \checkmark & -- & \checkmark & -- & \checkmark & BCD  \\
			{\cite{statis3}} & -- & -- & -- & \checkmark &  \checkmark & Interference-Aware Deployment \\
			{\cite{statis5}} & -- & -- & \checkmark & -- & \checkmark & BCD  \\
			\midrule
			Our Scheme & \checkmark & \checkmark & \checkmark & \checkmark & \checkmark & BCD+DRL  \\
			\bottomrule
		\end{tabular}
	}
	\label{Scheme_Comparison}
\end{table*}

   In conventional UAV-assisted communication systems, the primary objective is typically to maximize system throughput. However, next-generation UAV communication networks are expected to be user-centric. Under the premise of heterogeneous and personalized user demands, another crucial performance metric is the user demand satisfaction gain, commonly referred to as user satisfaction. This metric is determined by the extent to which diverse throughput requirements are fulfilled. Depending on the type of requested data and individual user preferences, users may perceive different levels of satisfaction even under identical throughput conditions~\cite{statis1,statis3,statis5,statis6}. 
   In \cite{statis1}, Tian \textit{et al.} jointly considered task processing delay and energy efficiency, constructed a novel user satisfaction model, and formulated a total satisfaction maximization problem to optimize both offloading decisions and UAV scheduling strategies. Similarly, Lai \textit{et al.}~\cite{statis3} proposed an interference-aware deployment algorithm to reduce co-channel interference among UAV-BSs, thereby improving user satisfaction and ensuring that most GUs meet their minimum data rate requirements. In~\cite{statis5}, Barick \textit{et al.} utilized the cooperative capabilities of UAVs and ground edge servers to maximize user satisfaction, ultimately enhancing the profit of the service provider. Moreover, Zhang \textit{et al.} ~\cite{statis6} developed a deep reinforcement learning (DRL)-based self-regulating approach to maximize the cumulative user satisfaction score during dynamic UAV network operations, such as UAV arrivals and departures. The distinctions between our proposed scheme and the relevant existing schemes are summarized in Table~\ref{Scheme_Comparison}.

	\subsection{Motivations and Contributions}
Based on the above studies, several critical challenges remain unsolved, including meeting personalized service demands of UDs, accurately analyzing and modeling ground obstacles of diverse shapes, achieving efficient 3D obstacle avoidance for UAVs, and effectively handling UAV battery depletion. In light of the aforementioned challenges, the main contributions of this study are summarized as follows:
\begin{itemize}
    \item We consider a heterogeneous communication scenario in a mixed-obstacle environment, where UAVs serve as aerial BSs to assist ground wireless communications. UAV trajectory planning is influenced by ground obstacles with diverse geometries. When UDs are located within obstacle-covered regions, UAVs must approach the obstacles to ensure stable service. To enhance system robustness, UAV failures due to battery depletion are also considered in the design of subsequent association decisions. Furthermore, UDs are classified into multiple urgency levels based on their service requirements, and a corresponding user satisfaction model is constructed. To optimize overall system performance, we jointly consider 3D trajectory planning, transmit beamforming, and association decisions between UAVs and UDs, with the objective of maximizing overall user satisfaction.
    
    \item To address the resulting high-dimensional and non-convex optimization problem, a block coordinate descent (BCD) framework is adopted to decouple it into two tractable subproblems. For the beamforming subproblem, a closed-form solution is derived based on a bisection-based water-filling algorithm. Given the complexity of the subproblem involving trajectory planning and UAV-UD association strategies, a DRL approach based on the proximal policy optimization (PPO) algorithm is proposed to efficiently learn the optimal control policy.
    
    \item  We examine the complexity of the proposed algorithm and conduct extensive simulation experiments. The results show that, in all considered scenarios, the proposed scheme outperforms the benchmark scheme in terms of user satisfaction, robustness, and computational complexity.

\end{itemize}
	\subsection{Organization and Notation}
    The remainder of this paper is organized as follows. Section II introduces the proposed UAV-assisted communication system model and formulates the optimization problem. Section III presents the proposed solution approach in detail. Section IV provides simulation results to validate the performance of the proposed algorithm. Finally, Section V concludes the paper.

    \textbf{Notation:} Scalars are denoted by italic letters \(x\), vectors by bold lowercase letters \(\mathbf{x}\), and matrices by bold uppercase letters \(\mathbf{X}\). The set of all complex-valued matrices of size $N \times M$ is denoted by $\mathbb{C}^{N \times M}$. For a complex vector $\mathbf{a}$, $\|\mathbf{a}\|$ denotes its Euclidean norm, $\mathbf{a}^\mathrm{T}$ denotes the transpose, and $\mathbf{a}^\mathrm{H}$ denotes the Hermitian (conjugate transpose). The notation \( \succeq \) denotes the matrix inequality under the positive semidefinite cone. The expectation operator is denoted by $\mathbb{E}[\cdot]$. The operator $\otimes$ represents the Kronecker product. The notation $[\cdot]_{k,k}$ refers to the $k$-th diagonal element of a matrix. The operator $[\cdot]^+ = \max(\cdot, 0)$ denotes the non-negative projection.

   	\section{System Model}
     In this section, we provide an overview of a low-altitude UAV-assisted communication scenario with mixed ground obstacles. We then present a detailed description of the scenario description, obstacle model, communication model, and user satisfaction model.
    \subsection{Scenario Description}
    \begin{figure}[t]
		\centering
		\includegraphics[width=\columnwidth]{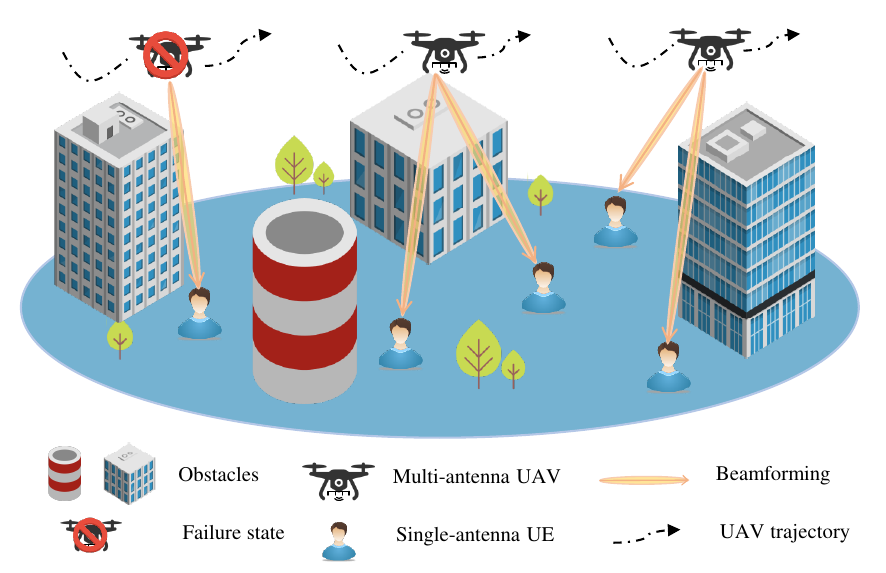}
		\caption{System model.}
		\label{fig:sys-model}
	\end{figure}
As shown in Fig.~\ref{fig:sys-model}, we consider a 3D UAV-assisted communication system operating in an environment with ground obstacles of various shapes. In this system, a set of $N$ UAVs, each equipped with $N_{\mathrm{r}}$ antennas, act as aerial BSs to provide wireless services for $K$ single-antenna UDs. The sets of UAV and UD indices are denoted by $\mathcal{N} = \{1, \dots, n,\dots N\}$ and $\mathcal{K} = \{1, \dots,k,\dots K\}$, respectively. The total mission duration $T$ is uniformly divided into $L$ non-overlapping intervals, each of duration $\delta_t =\frac{T}{L}$, to enable tractable modeling and optimization. The set of time slots is defined as $\mathcal{T} = \{1, \dots,t,\dots, T\}$. A 3D Cartesian coordinate system is adopted for system modeling. Specifically, the location of UAV $n$ at time slot $t$ is denoted by $\mathbf{q}_n[t] = \left( x_n[t], y_n[t], z_n[t] \right)^T$, while the location of UD $k$ is fixed, and given by $\mathbf{u}_k = \left( x_k, y_k, 0 \right)^T$. The UAV operates within the feasible region $\mathcal{B} = \{ (x,y,z) \in \mathbb{R}^3 \;\mid\; 0 \leq x \leq x_{\max}, \; 0 \leq y \leq y_{\max}  ,\; H_{\min} \leq z \leq H_{\max}\}$, where $x_{\max}$ and $y_{\max}$ represent the horizontal flight boundaries, and $H_{\min}$ and $H_{\max}$ denote the minimum and maximum allowable UAV altitudes, respectively.  Accordingly, the UAV flight positions must satisfy the following boundary constraints:
\begin{align}
\mathbf{q}_n[t] \in \mathcal{B}, \forall n \in \mathcal{N}.
\label{eq:x_constraint}
\end{align}

To ensure collision avoidance during communication tasks, a minimum safety distance $d_{\min}$ is enforced between any two UAVs. Moreover, to guarantee trajectory feasibility and control accuracy, each UAV is subject to a maximum velocity $v_{\max}$ and a maximum acceleration $a_{\max}$. Consequently, in each time slot, the UAV trajectories must satisfy the following constraints:
\begin{align}
\mathbf{q}_n[t+1] &= \mathbf{q}_n[t] + \mathbf{v}_n[t]\,\delta_t + \frac{1}{2} \mathbf{a}_n[t]\,\delta_t^2, 
 \label{eq:position_update} \\
\left\| \mathbf{q}_n[t] - \mathbf{q}_{n'}[t] \right\|^2 &\geq d_{\min}^2, 
\quad \forall n,n' \in \mathcal{N},\; n \ne n', 
 \label{eq:collision_avoidance} \\
\left\| \mathbf{v}_n[t] \right\| &\leq v_{\max}, 
\quad \forall t \in \mathcal{T}, 
 \label{eq:max_velocity} \\
\left\| \mathbf{a}_n[t] \right\| &\leq a_{\max}, 
\quad \forall t \in \mathcal{T}.
 \label{eq:max_acceleration}
\end{align}

The propulsion power consumption of UAV $n$ during flight at time slot $t$ is given by
\begin{align}
		P_n^{\text{fly}}[t] = & \frac{1}{2} d_0 \rho_0 g_0 A_1 \|\mathbf{v}_n[t]\|^3 
		+ P_0 \left(1 + \frac{3 \|\mathbf{v}_n[t]\|^2}{U_{\text{tip}}^2} \right) \notag \\
		& + P_1 \left(1 + \frac{\|\mathbf{v}_n[t]\|^4}{4v_0^4} - \frac{\|\mathbf{v}_n[t]\|^2}{2v_0^2} \right)^{\frac{1}{2}},
	\end{align}
where $P_0$ denotes the blade profile power, $P_1$ is the induced power during hovering, and $d_0$ is the drag coefficient. The parameters $\rho_0$ and $g$ represent the air density and rotor solidity, respectively. $U_{\text{tip}}$ denotes the rotor blade tip speed, $A_1$ is the rotor disk area, and $v_0$ represents the mean rotor speed. Accordingly, the flight energy consumption of UAV $n$ at time slot $t$ is calculated by
\begin{equation}
E_n^{\text{fly}}[t] = P_n^{\text{fly}}[t] \cdot \delta_t.
\end{equation}

\subsection{Obstacle Model}
	\begin{figure}[t]
		\centering
		\subfigure[Cylindrical obstacles.]{
			\includegraphics[width=0.20\textwidth]{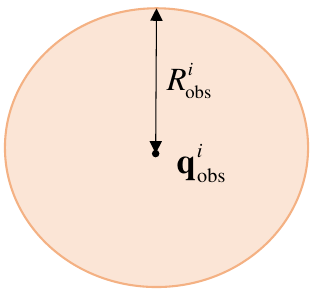}
		}
		\subfigure[Rectangular obstacle.]{
			\includegraphics[width=0.24\textwidth]{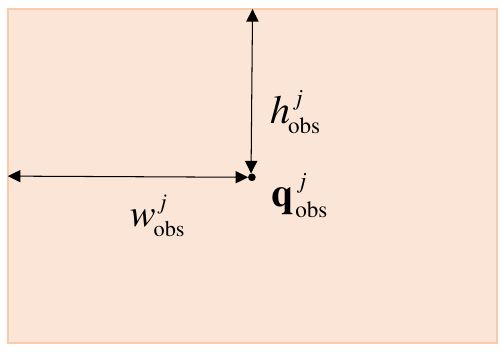}
		}
		\caption{The obstacle shape is projected on the x-y axis.}
		\label{fig:Obstacle shape}
	\end{figure}
In low-altitude environments, UAVs are prohibited from entering obstacle regions due to safety considerations. In other words, UAV trajectories must be constrained to avoid collisions with obstacles. In this work, we optimize the 3D trajectory of UAVs in the presence of ground obstacles with various shapes. Let the obstacle heights be denoted by a constant \( H_{\text{obs}} \), which satisfies the constraint \( H_{\min} < H_{\text{obs}} < H_{\max} \). When the UAV altitude satisfies $z_n[t] > H_{\text{obs}}$, it is permitted to fly over the obstacle, and the horizontal position constraints can be ignored. Conversely, if $z_n[t] \leq H_{\text{obs}}$, the UAV must satisfy the following horizontal constraints:

1) First, we consider a simplified scenario where cylindrical obstacles are projected onto the $x$-$y$ plane, as illustrated in Fig.~\ref{fig:Obstacle shape}(a). Assume that there are $I$ cylindrical obstacles, where the center and radius of obstacle $i$ are denoted by $\mathbf{q}_{\text{obs}}^i = \left( x_{\text{obs}}^i, y_{\text{obs}}^i \right)^\mathrm{T}$ and $R_{\text{obs}}^i$, respectively. The UAV trajectory constraint to avoid collision with these cylindrical obstacles is formulated as
\begin{equation}
\left\| \tilde{\mathbf{q}}_n[t] - \mathbf{q}_{\text{obs}}^i \right\|^2 \geq \left(R_{\text{obs}}^i\right)^2,\forall n \in \mathcal{N},\; i \in \mathcal{I},\label{eq:obs_circle}
\end{equation}
where $\tilde{\mathbf{q}}_n[t]$ denotes the UAV's horizontal position, defined as $ \tilde{\mathbf{q}}_n[t] = \left( x_n[t], y_n[t] \right)^T$.

2) To generalize the model, we also consider rectangular obstacles that are more common in practical environments, as shown in Fig. \ref{fig:Obstacle shape}(b). Suppose there are $J$ rectangular obstacles, with the center and side lengths of obstacle $j$ denoted by $\mathbf{q}_{\text{obs}}^j = \left( x_{\text{obs}}^j, y_{\text{obs}}^j \right)^T$ and $\mathbf{d}_{\text{obs}}^j = \left( w_{\text{obs}}^j, h_{\text{obs}}^j \right)^T$, respectively. Then, the UAV's trajectory constraint to avoid rectangular obstacles is given by
\begin{equation}
\left| \tilde{\mathbf{q}}_n[t] - \mathbf{q}_{\text{obs}}^j \right| \succeq {\mathbf{d}_{\text{obs}}^j}, \forall n \in \mathcal{N},\; j \in \mathcal{J}.\label{eq:obs_rect}
\end{equation}

\subsection{Communication Model}
To enable UAVs to better serve UDs, it is necessary to assign a reasonable UAV to each UD. We define the binary variable $\alpha_{n,k}[t] \in \{0,1\}$ to indicate the association between UAV $n$ and UD $k$. The UAV-UD association must satisfy the following constraints:
\begin{equation}
\sum_{n=1}^{N} \alpha_{n,k}[t] = 1, \quad \forall k \in \mathcal{K},\; t \in \mathcal{T},\label{eq:alpha1}
\end{equation}
\begin{equation}
\sum_{k=1}^{K} \alpha_{n,k}[t] \leq K_n, \quad \forall n \in \mathcal{N},\; t \in \mathcal{T},\label{eq:alpha2}
\end{equation}
constraint \eqref{eq:alpha1} ensures that each UD $k$ is associated with exactly one UAV, while constraint \eqref{eq:alpha2} ensures that each UAV $n$ is associated with at most $K_n$ UDs.

Specifically, when $E_n^{\text{res}}[t] = E_n^{\text{UAV}} - \sum_{s=1}^{t} E_n^{\text{fly}}[s] \leq 0$, the UAV is considered to be in a failure state, and we set $\beta_n = 0$; otherwise, $\beta_n = 1$. If UAV $n$ is available (i.e., $\beta_n = 1$) and provides communication service to UD $k$ at time slot $t$, then $\alpha_{n,k}[t] = 1$; in all other cases, $\alpha_{n,k}[t] = 0$. In particular, if UAV $n$ is initially assigned to UD $k$ but becomes unavailable due to energy depletion (i.e., $\beta_n = 0$), UD $k$ will be reassigned to another candidate UAV $n'$ to maintain communication and avoid service interruption. In this case, we set $\alpha_{n,k}[t] = 0$ and $\alpha_{n',k}[t] = 1$. The candidate UAV $n'$ must satisfy two conditions: it must be energy-sufficient, i.e., $E_{n'}^{\text{res}}[t] > 0$, and not overloaded, i.e., $\sum_{k=1}^{K} \alpha_{n',k}[t] \leq K_{n'}$.

Due to the complexity of the environment, the communication link between UAV $n$ and UD $k$ is often obstructed by trees and buildings, leading to Rician fading. Therefore, the channel between UAV $n$ and UD $k$ at time slot $t$ consists of a combination of line-of-sight (LoS) and non-line-of-sight (NLoS) components, and is expressed as
\begin{equation}
\mathbf{h}_{n,k}[t] = \sqrt{\rho d_{n,k}^{-\beta}[t]} \left( \sqrt{\frac{\varsigma}{1+\varsigma}} \overline{\mathbf{h}}_{n,k}^{\mathrm{LoS}}[t] + \sqrt{\frac{1}{1+\varsigma}} \tilde{\mathbf{h}}_{n,k}^{\mathrm{NLoS}}[t] \right),
\end{equation}
where $d_{n,k}[t] = \sqrt{\left\| \mathbf{q}_n[t] - \mathbf{u}_k \right\|^2}$ denotes the Euclidean distance between UAV $n$ and UD $k$ at time slot $t$, $\beta$ is the path loss exponent, and $\varsigma$ represents the Rician factor. The LoS component $\overline{\mathbf{h}}_{n,k}^{\mathrm{LoS}}[t] \in \mathbb{C}^{N_{\mathrm{r}} \times 1}$ is given by

\begin{align}
\overline{\mathbf{h}}_{n,k}^{\mathrm{LoS}}[t] =\ &\left( 
1,\, \ldots,\,
e^{-j\frac{2\pi b f_c}{c} \sin \varpi_{n,k}[t](a_x -1)\cos \theta_{n,k}[t]},
\right. \nonumber \\
&\left.
\ldots,\,
e^{-j\frac{2\pi b f_c}{c} \sin \varpi_{n,k}[t](A_x -1)\cos \theta_{n,k}[t]} 
\right) \nonumber \\
&\otimes \left(
1,\, \ldots,\,
e^{-j\frac{2\pi b f_c}{c} \sin \varpi_{n,k}[t](a_y -1)\cos \theta_{n,k}[t]},
\right. \nonumber \\
&\left.
\ldots,\,
e^{-j\frac{2\pi b f_c}{c} \sin \varpi_{n,k}[t](A_y -1)\cos \theta_{n,k}[t]} 
\right),
\end{align}
where $b$ denotes the antenna spacing factor, and $c=\mathbf{v}_n[t]$ is the UAV flying speed. The parameter $f_c$ represents the carrier frequency of the information signal. Let $a_x$ and $a_y$ denote the row and column indices of the uniform planar array,  respectively. The horizontal and vertical angles of departure (AoDs) from UD $k$ are defined as $\theta_{n,k}[t]$ and $\varpi_{n,k}[t]$, respectively. Specifically, the AoDs are given by
\begin{align}
\theta_{n,k}[t] &= \arccos \left( \frac{y_{n}[t] - y_k}{\left\| \mathbf{q}_n[t] - \mathbf{u}_k] \right\|} \right),\\
\varpi_{n,k}[t] &= \arcsin \left( \frac{z_n[t]}{\sqrt{ \left\| \mathbf{q}_n[t] - \mathbf{u}_k \right\|^2 }} \right).
\end{align}

Besides, the $\tilde{\mathbf{h}}_{n,k}^{\mathrm{NLoS}}[t] \in \mathbb{C}^{N_{\mathrm{r}} \times 1}$ denotes the NLoS component of the channel at time slot $t$. It is modeled as a zero-mean complex Gaussian random vector with independent and identically distributed elements and unit variance, i.e., $\tilde{\mathbf{h}}_{n,k}^{\mathrm{NLoS}}[t] \sim \mathcal{CN}(\mathbf{0}, \mathbf{I})$. Accordingly, the received signal at UD $k$ from UAV $n$ at time slot $t$ can be expressed as
\begin{align}
y_{n,k}[t] &= \mathbf{w}_{n,k}^H[t] \mathbf{h}_{n,k}[t] s_{n,k}[t] \nonumber \\
&\quad + \sum_{j=1}^{N} \sum_{\substack{i=1 \\ i \ne k}}^{K} 
\alpha_{j,i}[t] \mathbf{w}_{n,k}^H[t] \mathbf{h}_{j,i}[t] s_{j,i}[t] \nonumber \\
&\quad + \mathbf{w}_{n,k}^H[t] \mathbf{n},
\label{eq:signal} 
\end{align}
where $\mathbf{w}_{n,k}[t] \in \mathbb{C}^{N_{\mathrm{r}} \times 1}$ is the unit-norm beamforming vector between UAV $n$ and UD $k$ at time slot $t$, satisfying $\mathbf{w}_{n,k}^\mathrm{H}[t] \mathbf{w}_{n,k}[t] = 1$. The symbol $s_{n,k}[t]$ denotes the signal transmitted from UAV $n$ to UD $k$, with $\mathbb{E} \left[ |s_{n,k}[t]|^2 \right] = 1$. In addition, $\mathbf{n} \sim \mathcal{CN}(0, \sigma^2 \mathbf{I})$ denotes the additive white Gaussian noise vector with zero mean and covariance $\sigma^2 \mathbf{I}$. Hence, the signal-to-interference-plus-noise ratio is given by

\begin{align}
\Upsilon_{n,k}[t] = \frac{\left| \mathbf{w}_{n,k}^H[t] \mathbf{h}_{n,k}[t] \right|^2}
{\sum_{j=1}^{N} \sum_{\substack{i=1 \\ i \ne k}}^{K} \alpha_{j,i}[t] \left| \mathbf{w}_{n,k}^H[t] \mathbf{h}_{j,i}[t] \right|^2 + \sigma^2}.
\end{align}

According to Shannon's capacity formula, the achievable data rate from UAV $n$ to UD $k$ is
\begin{align}
R_{n,k}[t] = B_0 \log_2 \left(1 + \Upsilon_{n,k}[t] \right),
\end{align}
where $B_0$ denotes the channel bandwidth.
\subsection{Satisfaction Model}
Let \( S_{n,k}[t] \in [0,1] \) denote the satisfaction level of UD \(k\) with respect to the communication service provided by UAV \(n\) at time slot \(t\). To capture the personalized service demands of UDs, the communication requirement of UD \(k\) is modeled using a generalized satisfaction mapping function, as described in~\cite{statis2}
\begin{equation}
f_{n,k}[t] = \frac{1}{1 + \exp\left( -\lambda_{n,k}[t] \left( R_{n,k}[t] -  R_{n,k}^{\text{target}} + \frac{C}{\lambda_{n,k}[t]} \right) \right)},
\end{equation}
where \( \lambda_{n,k}[t] \) represents the dynamic urgency level of the transmission task between UAV \(n\) and UD \(k\). A higher value of \( \lambda_{n,k}[t] \) indicates a more urgent task, which requires a higher transmission rate. The term \( R_{n,k}^{\text{target}} \) denotes the corresponding required transmission rate, and \( C > 7 \) is a predefined shaping constant. When the actual data rate exceeds the target rate, i.e., \( R_{n,k}[t] >  R_{n,k}^{\text{target}} \), the output of the satisfaction function satisfies \( f_{n,k} > \frac{1}{1 + \exp(-C)} \approx 1 \). The satisfaction indicator \( S_{n,k}[t] \) is then defined as

\begin{equation}
S_{n,k}[t] =
\begin{cases}
f_{n,k}[t], & \text{if } \alpha_{n,k}[t] = 1, \\
0, & \text{if } \alpha_{n,k}[t] = 0,
\end{cases}
\end{equation}
when there is no association between UD \(k\) and UAV \(n\), that is, \( \alpha_{n,k}[t] = 0 \), then the satisfaction level becomes \( S_{n,k}[t] = 0 \).
\subsection{Problem Formulation}
Our objective is to maximize the overall user satisfaction by jointly optimizing the UAV flying trajectory $\mathbf{Q} \triangleq \{\mathbf{q}_{n}[t], \forall n \in \mathcal{N}, t \in \mathcal{T}\}$, transmit beamforming $\mathbf{W} \triangleq \{\mathbf{w}_{n,k}[t], \forall n \in \mathcal{N}, k \in \mathcal{K},t \in \mathcal{T}\}$, and the binary association variable between UDs and UAVs $\bm{\alpha} \triangleq \{\alpha_{n,k}[t], \forall n \in \mathcal{N}, k \in \mathcal{K} , t \in \mathcal{T}\}$. The corresponding optimization problem is formulated as

	\begin{subequations} 
		\begin{align}     
			\max_{\mathbf{Q}, \mathbf{W}, \boldsymbol{\alpha}} \quad & S_{\text{total}} = \sum_{t=1}^{T} \sum_{k=1}^{K} S_n[t] = \sum_{t=1}^{T} \sum_{n=1}^{N} \sum_{k=1}^{K} S_{n,k}[t] \label{eq:a} \\   
			\text{s.t.} & \quad \eqref{eq:x_constraint},\eqref{eq:collision_avoidance}, \eqref{eq:max_velocity},\eqref{eq:max_acceleration}, \eqref{eq:obs_circle},\eqref{eq:obs_rect}
            ,\eqref{eq:alpha1},\eqref{eq:alpha2}, \label{eq:b} \\        
			& \quad \text{Tr}\left( \mathbf{W}_{n}^{H}[t] \mathbf{W}_{n}[t] \right) \leq P_n^{\max},  n \in \mathcal{N}, \label{eq:d} \\
            & \quad S_n[t] \leq K_n, S_{\text{total}} \leq K,\forall n \in \mathcal{N}, \label{eq:f}
		\end{align} 
		\label{eq:sum} 
	\end{subequations} 
where \( K_n \) denotes the maximum number of UDs associated with UAV \(n\). Constraint~\eqref{eq:x_constraint} defines the UAV flight boundary limits in the 3D space. Constraint~\eqref{eq:collision_avoidance} ensures a minimum safety distance between any two UAVs. Constraints~\eqref{eq:max_velocity} and \eqref{eq:max_acceleration} impose limits on UAV velocity and acceleration, respectively. Constraints~\eqref{eq:obs_circle} and \eqref{eq:obs_rect} restrict the UAV trajectories to avoid cylindrical and rectangular obstacles, respectively. Constraints~\eqref{eq:alpha1} and \eqref{eq:alpha2} govern the binary association decisions between UAVs and UDs. Constraint~\eqref{eq:d} ensures that the transmit power of UAV $n$ does not exceed its maximum allowable value, where $\mathbf{W}_{n}[t]$ represents the beamforming vector at time slot $t$. Constraint~\eqref{eq:f} enforces the user satisfaction requirement.

	\section{Problem Solution}\label{pro:s}
It is noteworthy that problem~\eqref{eq:sum} is a non-convex optimization problem involving disjunctive planning. In general, such non-convex problems are difficult to solve directly due to their inherent computational complexity. Our analysis reveals that the main challenges arise from the non-convexity of both the objective function and several constraints, particularly the trajectory-related restrictions imposed by obstacles, as described in constraints~\eqref{eq:obs_circle} and~\eqref{eq:obs_rect}. To address these challenges, we propose a low-complexity heuristic algorithm based on the BCD framework. The original problem is decomposed into two subproblems: a beamforming subproblem and a joint trajectory and association subproblem. For the beamforming subproblem, given the UAV trajectory $\mathbf{Q}$ and UAV-UD association variables $\boldsymbol{\alpha}$, we design a bisection-based water-filling algorithm that yields a closed-form solution for the beamforming vectors. For the joint trajectory and association subproblem, given the beamforming $\mathbf{W}$, we formulate the problem as a Markov decision process (MDP) and solve it using the PPO algorithm. Through alternating optimization of these two algorithms, the system is guided to converge more rapidly and robustly toward a better solution. 

Furthermore, the proposed framework remains applicable even when faced with various real-world disruptions. For example, GPS drift can cause trajectory deviations, which may be mitigated using robust localization or sensor fusion techniques. Additionally, the energy constraints of UAVs, which are influenced by wind disturbances, can be alleviated through adaptive energy management strategies or robust trajectory re-planning. These modifications primarily expand the state/action spaces and constraints but do not alter the core structure of the proposed framework. The details of the proposed solution framework are presented in the following section.
\subsection{Beamforming Optimization}
In our model, the wireless channel is characterized by a Rician fading distribution, where the presence or absence of a LoS path is implicitly determined by the surrounding environment. Since the beamforming vectors are optimized based on channel state information (CSI), which inherently reflects the impact of obstacles, the optimization process naturally avoids directions with severe attenuation. Consequently, we do not explicitly incorporate geometric obstacle constraints in the beamforming optimization. 

Considering that the beamforming matrices of UAVs are independent across different time slots, we omit the time index for simplicity in the beamforming subproblem. Given the UAV positions \( \mathbf{Q} \) and UAV-UD association indicators \( \boldsymbol{\alpha} \), the beamforming optimization can be reformulated as follows:
\begin{subequations} 
\begin{align}
\max_{ \mathbf{W}} \quad & S_{\text{total}} \label{eq:a1} \\ 
\text{s.t.} \quad & \text{Tr} \left( \mathbf{W}_{n}^{H} \mathbf{W}_{n} \right) \leq P_n^{\max}, \forall n \in \mathcal{N}. \label{eq:b1} 
\end{align}
\end{subequations} 

This problem is equivalent to the following sum-rate maximization problem:
\begin{subequations}
\begin{align}
\max_{\mathbf{W}} \quad & 
\sum_{n=1}^{N} \sum_{k=1}^{K} 
R_{n,k}[t]  \label{eq:a2} \\ 
\text{s.t.} \quad & \text{Tr} \left( \mathbf{W}_{n}^{H} \mathbf{W}_{n} \right) \leq P_n^{\max}, \forall n \in \mathcal{N}. 
\label{eq:b2} 
\end{align}
\label{eq:P2} 
\end{subequations}
To eliminate the interference among UDs served by the same UAV in communication system, the zero-forcing (ZF) beamforming strategy is adopted.  Specifically, for each UAV $n$, the beamforming vector $\mathbf{w}_{n,k}[t]$ is designed such that $\mathbf{w}_{n,k}^H[t] \mathbf{h}_{n,j}[t] = 0,  \forall j \neq k$, which ensures that the signal intended for UD $k$ is orthogonal to the channels of other UDs served by the same UAV. The ZF strategy is feasible under the condition $N_{\mathrm{r}} \geq K_n$, under which it effectively suppresses inter-UD interference with tractable complexity and near-optimal performance. The received signal of UAV \(n\), as defined in \eqref{eq:signal}, can be rewritten as $\mathbf{y}_n = \mathbf{W}_n^H \mathbf{H}_n \mathbf{s}_n + \mathbf{W}_n^H \mathbf{n}$, where \( \mathbf{y}_n = [y_{n,1}, \ldots, y_{n,K}]^T \), \( \mathbf{s_n} = [s_{n,1}, \ldots, s_{n,K}]^T \), and \( \mathbf{H}_n \in \mathbb{C}^{K \times N_{\mathrm{r}}} \) is the channel matrix from UAV \(n\) to the associated UDs, with the \(k\)-th row denoted by \( \mathbf{h}_{n,k}^T \). The ZF beamforming matrix is defined as
\begin{equation}
\mathbf{W}_n = \mathbf{H}_n^H \left( \mathbf{H}_n \mathbf{H}_n^H \right)^{-1} \mathbf{P}_n^{\frac{1}{2}} = \widetilde{\mathbf{H}}_n \mathbf{P}_n^{\frac{1}{2}},
\label{eq:zf_beamforming}
\end{equation}
where \( \widetilde{\mathbf{H}}_n = \mathbf{H}_n^H \left( \mathbf{H}_n \mathbf{H}_n^H \right)^{-1} \), and the power allocation matrix \( \mathbf{P}_n = \text{diag}(p_{n,1}, \ldots, p_{n,K}) \). According to the water-filling algorithm, the optimal power allocation is given by
\begin{equation}
p_{n,k} = \left[ \frac{1}{\mu_n} - \sigma^2 \cdot \left[ \left( \widetilde{\mathbf{H}}_n^H \widetilde{\mathbf{H}}_n  \right)^{-1} \right]_{k,k} \right]^+,
\label{eq:power_allpcation}
\end{equation}
where \( \mu_n \) is a normalization factor to ensure the power constraint $\sum_{k=1}^{K} p_{n,k} - P_n^{\max}\leq \varepsilon$. The parameter \( \varepsilon \) denotes the desired accuracy threshold. The algorithm can be summarized in Algorithm~\ref{alg:water-filling}.

\begin{algorithm}[t]
\caption{Water-Filling and Bisection-Based Algorithm for Solving \eqref{eq:zf_beamforming}}
\label{alg:water-filling}
\begin{algorithmic}[1]
\STATE \textbf{Input:} $\mathbf{h}_{n,k}$, $\mathbf{H}_{n}$,  $\sigma^2$, $\varepsilon$.
\STATE \textbf{Output:} Optimal beamforming based on~\eqref{eq:power_allpcation}.
\STATE Calculate matrix $\widetilde{\mathbf{H}}_n^H \widetilde{\mathbf{H}}_n$, and obtain $\mathbf{h}_{n,k}$.
\STATE Initialize $\mu_{\max} = \mu_{\min} = \mu_{\text{mid}}$.
\FOR{$n \leq N$}
    \STATE \textbf{Find upper and lower bounds for $\mu_n$:}
    \FOR{$k \leq K$}
        \STATE \textbf{if} $\mathbf{h}_{n,k} \sigma^2 \leq 1/\mu_{\max}$ \textbf{then} set $\mu_{\max} = 1 / (\mathbf{h}_{n,k} \sigma^2)$,
        \STATE \textbf{else} set $\mu_{\min} = 1 / (\mathbf{h}_{n,k} \sigma^2)$.
    \ENDFOR
    \STATE \textbf{Find the optimal $\mu_n$ using bisection method:}
    \REPEAT
        \STATE Compute midpoint $\mu_{\text{mid}} = (\mu_{\max} + \mu_{\min}) / 2$.
        \STATE Calculate the sum of power allocations $P_{\text{sum}} = \sum_{k=1}^{K} \left[ \dfrac{1}{\mu_n} - \sigma^2 \cdot \left[ \left( \widetilde{\mathbf{H}}_n^H \widetilde{\mathbf{H}}_n \right)^{-1} \right]_{k,k} \right]^+$.
        \STATE \textbf{if} $P_{\text{sum}} > P_n^{\max}$ \textbf{then} set $\mu_{\min} = \mu_{\text{mid}}$,
        \STATE \textbf{else} set $\mu_{\max} = \mu_{\text{mid}}$.
    \UNTIL{ $| P_{\text{sum}} - P_n^{\max}| \leq \varepsilon$}
\ENDFOR
\end{algorithmic}
\end{algorithm}

\subsection{Trajectory and Association Optimization}
After obtaining the beamforming matrix \(\mathbf{W}\) from Algorithm~\ref{alg:water-filling}, we proceed to optimize the UAV trajectory \(\mathbf{Q}\) and UAV-UD association decisions \(\bm{\alpha}\). Given \(\mathbf{W}\), the original problem \eqref{eq:sum} can be simplified as
	\begin{subequations} 
		\begin{align}     
			\max_{\mathbf{Q}, \boldsymbol{\alpha}} \quad & S_{\text{total}} \label{eq:a} \\   
			\text{s.t.} & \quad \eqref{eq:x_constraint},\eqref{eq:collision_avoidance}, \eqref{eq:max_velocity},\eqref{eq:max_acceleration}, \eqref{eq:obs_circle},\eqref{eq:obs_rect}
            ,\eqref{eq:alpha1},\eqref{eq:alpha2}, \label{eq:b} \\          
            & \quad S_n[t] \leq K_n, S_{\text{total}} \leq K,\forall n \in \mathcal{N}. \label{eq:f2}
		\end{align} 
		\label{P2} 
	\end{subequations} 
Since problem~\eqref{P2} is a non-convex integer programming problem, it presents significant computational challenges when approached directly. In addition, the inherent uncertainty and variability introduced by time-varying channel conditions and heterogeneous task characteristics further hinder the applicability of traditional offline optimization techniques, which often assume static or fully known environments. To overcome these limitations and enable real-time, adaptive decision-making for heterogeneous resource allocation, we introduce a DRL framework to derive the optimal joint configuration strategy. 
\subsubsection{Modeling of MDP}
To facilitate the application of DRL techniques, we reformulate the original optimization problem~\eqref{P2} as an MDP. In the context of DRL, an MDP is commonly characterized by a tuple \((\mathcal{S}, \mathcal{A}, \mathcal{R}, \gamma)\), where \(\mathcal{S}\) denotes the state space, \(\mathcal{A}\) represents the action space, \(\mathcal{R}\) specifies the reward function, and \(\gamma \in [0, 1]\) is the discount. The discount factor \(\gamma\) reflects the agent’s preference for long-term returns, with values closer to 1 indicating a greater emphasis on future rewards. The definitions of the state space, action space, and reward function are provided as follows.

\textit{State:} The state is the basis of decision-making. We define the state space within each time slot \( t \) as
\begin{equation}
s_t = \left\{ \mathbf{q}_n[t], E_n^{\text{res}}[t], \mathbf{v}_n[t], \lambda_{n,k}[t], \forall n,k \right\},
\end{equation}
where \(\mathbf{q}_n[t]\) denotes the position of UAV \(n\), \(E_n^{\text{res}}[t]\) represents its residual battery level, and \(\mathbf{v}_n[t] = [\|\mathbf{v}_n[t]\|, \varphi_n[t]]\) characterizes the UAV's flight velocity, including its magnitude and direction. In addition, $\lambda_{n,k}[t]$ represents the urgency level of UAV \(n\) associated with UD \(k\).

\textit{Action:} Since the optimal beamforming \(\mathbf{W}\) solution has already been determined, the action space thus includes
\begin{equation}
a_t = \left\{ \mathbf{a}_n[t], \alpha_{n,k}[t],\forall n,k \right\},
\end{equation}
where $\mathbf{a}_n[t] = [\|\mathbf{a}_n[t]\|, \phi_n[t]]$ denotes the magnitude and direction of the UAV acceleration, and \(\alpha_{n,k}[t]\) represents the association indicator between UAV \(n\) and UD \(k\).

\textit{Reward:} The reward function considered in this paper consists of two components: reward and penalty. The definition of the penalty term is given as follows:

\begin{equation}
\begin{aligned}
r_t = 
\begin{cases}
S_{\text{total}} - \kappa_{1} \sum_{n=1}^{N} d^{2}(\mathbf{q}_{n}[t], \mathcal{B}),
\\
\qquad\text{if } \mathbf{q}_n[t] \notin \mathcal{B}, \\[6pt]
S_{\text{total}} - \kappa_{2} \sum_{i=1}^{I} \sum_{n=1}^{N} ( D_{ni}^c)^2,
\\
\qquad\text{if } 
\|\tilde{\mathbf{q}}_n[t] - \mathbf{q}_{\text{obs}}^i\| \le R_{\text{obs}}^i,
~ z_n[t] \le H_{\text{obs}}, \\[6pt]
S_{\text{total}} - \kappa_{3} \sum_{j=1}^{J} \sum_{n=1}^{N} (D_{nj}^r)^2,
\\
\qquad\text{if }
\left| \tilde{\mathbf{q}}_n[t] - \mathbf{q}_{\text{obs}}^j \right| 
\preceq \mathbf{d}_{\text{obs}}^j,
~ z_n[t] \le H_{\text{obs}}, \\[6pt]
S_{\text{total}} - \kappa_{4} 
\sum_{n<n'} \exp\!\left(
-\frac{\|\mathbf{q}_n[t] - \mathbf{q}_{n'}[t]\|}{d_{\min}}
\right),
\\
\qquad\text{if }
\|\mathbf{q}_n[t] - \mathbf{q}_{n'}[t]\|^2 \le d_{\min}^2,
\end{cases}
\end{aligned}
\end{equation}
where  \(D_{ni}^c=R_{\text{obs}}^i- \|\tilde{\mathbf{q}}_n[t] - \mathbf{q}_{\text{obs}}^i\|  \) and \(D_{nj}^r=\max( {\mathbf{d}_{\text{obs}}^j}-\left| \tilde{\mathbf{q}}_n[t] - \mathbf{q}_{\text{obs}}^j \right| ) \), respectively. 

Specifically, when the UD is located inside the obstacle, a reward is given to encourage the UAV to approach the obstacle. The definition of the reward term is as follows:
\begin{equation}
r_t =
\begin{cases}
S_{\text{total}} 
+ \sum_{i=1}^{N} \sum_{k=1}^{K}
\exp\!\left(\kappa_{5}\left(d_{\text{NFZ}} - D_{ni}^c\right)\right),
\\
\qquad\text{if } 
R_{\text{obs}}^i < \|\tilde{\mathbf{q}}_n[t] - \mathbf{q}_{\text{obs}}^i\|
\le \left(R_{\text{obs}}^i + d_{\text{safe}}^c\right),
\\[8pt]
S_{\text{total}} 
+ \sum_{i=1}^{N} \sum_{k=1}^{K}
\exp\!\left(\kappa_{6}\left(d_{\text{NFZ}} - D_{nj}^r\right)\right),
\\
\qquad\text{if }
\mathbf{d}_{\text{obs}}^j 
\prec \left| \tilde{\mathbf{q}}_n[t] - \mathbf{q}_{\text{obs}}^j \right|
\preceq \left(\mathbf{d}_{\text{obs}}^j + \mathbf{d}_{\text{safe}}^r\right),
\end{cases}
\end{equation}
where $d_{\text{safe}}^c$ denotes the safety margin for the cylindrical obstacle, and $\mathbf{d}_{\text{safe}}^r$ represents the corresponding direction vector from the UAV to the Rectangular obstacle used for collision avoidance. Additionally, $\kappa_1$, $\kappa_2$, $\kappa_3$, $\kappa_4$ ,$\kappa_5$ and $\kappa_6$ represent the coefficients for the penalty and reward terms, and they should be carefully adjusted to ensure convergence.

\subsubsection{PPO-Based Algorithm}
\begin{figure*}[t]
		\centering
		\includegraphics[width=1.6\columnwidth]{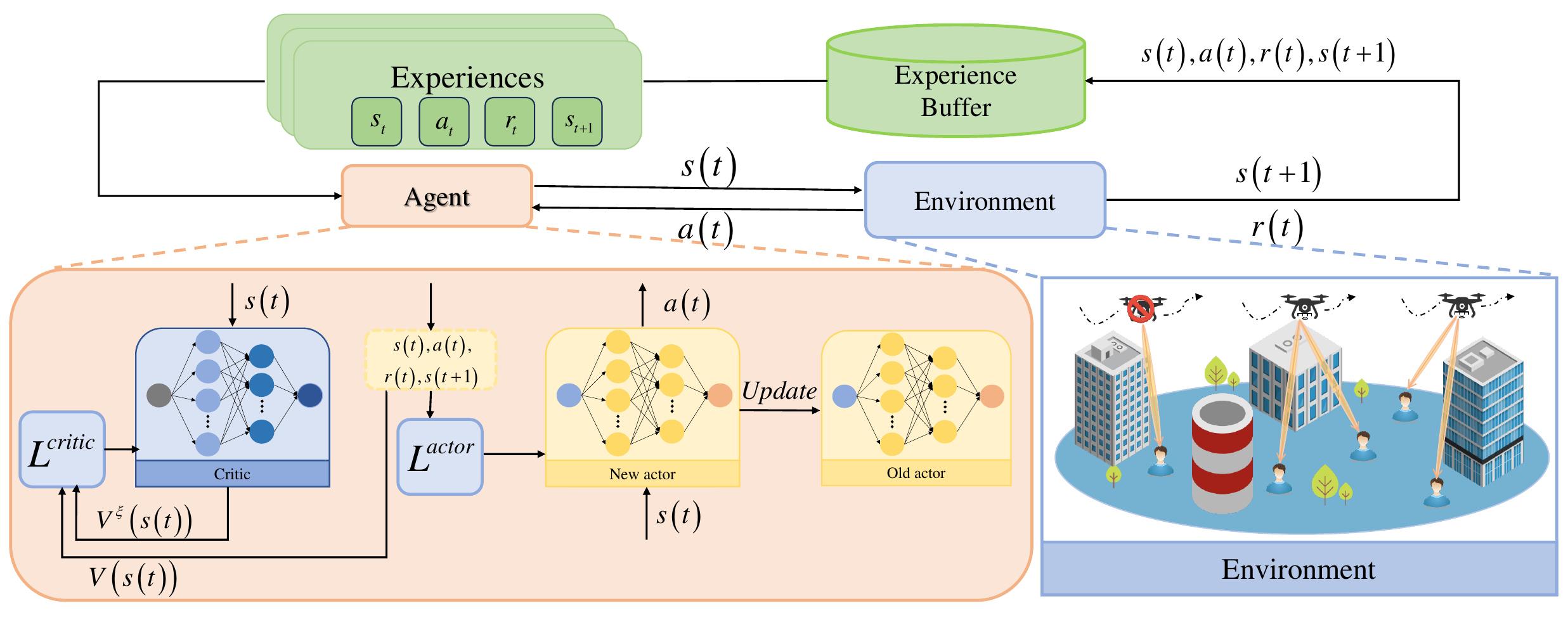}
		\caption{The PPO-based framework.}
		\label{fig:PPO}
	\end{figure*}
In this section, we present the proposed PPO algorithm in detail. Given the presence of both continuous and discrete values in the action and state spaces, explicitly modeling the environment becomes intractable. To address this challenge, we employ the PPO algorithm, which enables approximation of the optimal policy without requiring an explicit representation of the state and action spaces. The DRL training framework based on PPO is depicted in Fig.~\ref{fig:PPO}. In contrast to conventional policy gradient methods, PPO incorporates a clipped surrogate objective that restricts the magnitude of policy updates, thereby enhancing training stability and preventing performance collapse due to overly aggressive updates. Under this formulation, the learning objective of the agent can be expressed as
\begin{equation}
    \max_{\pi} \sum_{t=1}^{T} \mathbb{E}_{(s_t,a_t)\sim \rho_\pi}
\bigl[\gamma^{\,t-1} r_t \bigr].
\end{equation}

In addition, PPO incorporates the generalized advantage estimation (GAE) technique to strike a balance between bias and variance, thereby reducing the number of samples required for effective policy learning. The advantage function \(\hat{A}_t\), which quantifies how much better an action is compared to the expected return under the current policy, is defined as
\begin{equation}
    \hat{A}_t = \delta_t + (\gamma \lambda) \delta_{t+1} + (\gamma \lambda)^2 \delta_{t+2} + \cdots + (\gamma \lambda)^{T - t} \delta_T,
\end{equation}
where the temporal difference error $\delta_t$ is given by $\delta_t = r_t + \gamma V(s_{t+1}) - V(s_t)$, in which \(\gamma\) denotes the discount factor, and \(\lambda\) is the GAE coefficient that controls the trade-off between bias and variance in the advantage estimation. This paper adopts a PPO algorithm based on the clip mechanism to train the actor–critic networks. The probability ratio between the new and old policy is defined as $r_t(\theta) = \frac{\pi_\theta(a_t \mid s_t)}{\pi_{\theta_{\text{old}}}(a_t \mid s_t)}$, where $\theta$ and $\theta_{\text{old}}$ denote the parameters of the current and previous actor networks, respectively. Accordingly, the loss function for the actor network is formulated as
\begin{equation}
L^{\text{actor}} = \mathbb{E}_t \left[ \min \left( r_t(\theta) \hat{A}_t, \; \text{clip} \left( r_t(\theta), 1 - \epsilon, 1 + \epsilon \right) \hat{A}_t \right) \right],
\label{eq:loss_actor}
\end{equation}
where \(\mathbb{E}_t[\cdot]\) denotes the empirical expectation over a batch of sampled timesteps, and \(\text{clip}(\cdot)\) refers to the clipping function that constrains the range of the policy ratio \(r_t(\theta)\) to prevent excessively large updates. The parameter \(\epsilon\) is a predefined hyperparameter that determines the allowable deviation range for the clipped objective. In practice, the critic network is optimized by minimizing a value-based loss to ensure stable policy evaluation. By incorporating the mean squared error between the predicted state value and the empirical return, the loss function of the critic network is defined as
\begin{equation}
L^{\text{critic}}(\xi) = \mathbb{E}_t \left[ \left( V^{\xi}(s_{t+1}) - V(s_t)\right) ^2 \right],
\label{eq:loss_critic}
\end{equation}
where \(V^\xi(\cdot)\) denotes the state-value function estimated by the critic network, and \(\xi\) represents the set of parameters associated with the value network. Consequently, the actor and critic networks are updated periodically by computing the gradients of the loss functions defined in Equations~\eqref{eq:loss_actor} and~\eqref{eq:loss_critic}, and applying them in an alternating optimization manner. In the environment considered in this work, the observed states are fed into the internal executor of the PPO framework, which determines the optimal actions to solve the underlying joint optimization problem in a model-free manner. The detailed training procedure of the proposed PPO-based learning framework is outlined in Algorithm~\ref{alg:ppo}.

\subsubsection{Complexity Analysis} The water-filling algorithm adopted in Algorithm~\ref{alg:water-filling} is a bisection-based iterative method that adjusts the dual variable to satisfy the total power constraint with a predefined accuracy \(\varepsilon\). In each iteration, the algorithm computes the inverse of the effective channel matrix, which incurs a computational complexity of \(O(N_{\mathrm{r}}^3)\), and subsequently updates the power allocation for \(K\) UDs. Given that the number of iterations required to achieve the target precision is \(O\left(\log_2\left(\frac{1}{\varepsilon}\right)\right)\), the overall time complexity of the water-filling procedure can be expressed as $O\left(K \cdot \log_2\left(\frac{1}{\varepsilon}\right) \cdot N_{\mathrm{r}}^3\right)$. In the PPO algorithm adopt in Algorithm \ref{alg:ppo}, both the actor and critic networks are implemented as multi-layer perceptrons. For the \(s\)-th hidden layer, the computational complexity is given by $O(L_{s-1} L_s + L_s L_{s+1})$, where \(L_s\) denotes the number of neurons in layer \(s\). Consequently, the total complexity for a network with \(S\) layers is $O\left(\sum_{s=2}^{S-1} \left( L_{s-1} L_s + L_s L_{s+1} \right)\right)$. In our proposed framework, the overall computational complexity becomes
\begin{align}
O\Big( e^{\text{MAX}} \cdot \text{EL} \cdot
\big(&\sum_{s=2}^{S-1} ( L_{s-1} L_s + L_s L_{s+1} )  \nonumber \\
&+ K \cdot \log_2\!\left(\frac{1}{\varepsilon}\right) \cdot N_{\mathrm{r}}^3 \big)
\Big),
\end{align}
where \(\text{EL}\) denotes the  episode length and \(e^{\text{MAX}}\) denotes the maximum number of training episodes.
\begin{algorithm}[t]
\caption{PPO-based DRL Training Algorithm}
\label{alg:ppo}
\begin{algorithmic}[1]
\STATE \textbf{Input:} Network parameters of actor $\theta$, critic $\omega$, replay buffer $\mathcal{D}$, learning rate, discount factor $\gamma$.
\STATE \textbf{Output:} Optimal policy $\pi_{\theta}$.
\STATE Initialize environment.
\STATE Initialize critic networks $\omega$ and actor network $\phi$.
\STATE Initialize replay buffer: $\mathcal{D} = \emptyset$.
\STATE Initialize hyperparameters: learning rate, discount factor $\gamma$.
\FOR{each episode}
    \FOR{each environment step}
        \STATE Select action $a_t$ according to current policy: $a_t \sim \pi_{\theta}(\cdot | s_t)$;
        \STATE Execute action $a_t$, apply Algorithm~\ref{alg:water-filling} to obtain the optimal UAV beamforming matrix;
        \STATE Transmit to next state $s_{t+1}$, calculate reward $r_t$, store transition $\{s_t, a_t, r_t, s_{t+1}\}$ in $\mathcal{D}$.
    \ENDFOR
    \STATE Update actor network $\theta$ using objective \eqref{eq:loss_actor}.
    \STATE Update critic network $\omega$ using loss \eqref{eq:loss_critic}.
    \STATE Store policy entropy and log-probability in the replay buffer.
\ENDFOR
\end{algorithmic}
\end{algorithm}

	\section{SIMULATION RESULTS}\label{pro:s}
    In this section, we evaluate the performance of the proposed algorithm for multi-UAV communication systems from the perspective of user satisfaction. We analyze the impact of various factors on system performance, including the number and altitude of obstacles, UAV flight trajectories, and other relevant parameters. To evaluate the performance of the proposed algorithm, the following baseline algorithms are employed for comparison:

\begin{itemize}
    \item \textbf{Pure-PPO}: The UAVs are controlled by a PPO-based learning agent without incorporating any prior knowledge or beamforming optimization~\cite{PPO}. The beamforming vectors are directly embedded in the action space of the PPO agent and are jointly optimized along with UAV trajectories and UAV-UD association decisions.
  \item  \textbf{DDPG}: An off-policy actor–critic algorithm designed for continuous action spaces~\cite{DDPG}. It employs a deterministic policy to map states to actions via an actor network, while a critic network estimates the action-value function to guide policy improvement. DDPG enhances training stability and sample efficiency by leveraging experience replay and target networks with soft updates. During training, temporally correlated noise is added to promote exploration. Due to its suitability for high-dimensional, continuous control tasks, DDPG is widely applied in UAV trajectory and resource allocation problems.
    \item \textbf{Fixed UAV Scheme}: UAVs are statically deployed at predetermined positions, and no trajectory adaptation is performed during the operation.
    \item \textbf{Equal Beamforming Scheme}: The beamforming vector is designed to uniformly distribute the transmission power across all antenna elements, disregarding the CSI.
\end{itemize}

    \begin{table}[htbp]
\renewcommand{\arraystretch}{1.2}
\centering
\caption{System Parameters}
\label{tab:parameters}
\begin{tabular}{|c|c|}
\hline
Time period $T$ & $200\ \mathrm{s}$ \\
\hline
Interruption factor $\varsigma$ & $10$ \\
\hline
Time slot $\delta_t$ & $1\ \mathrm{s}$ \\
\hline
Channel power gain $\rho$ & $-30\ \mathrm{dB}$ \\
\hline
Noise power $\sigma$ & $-65\ \mathrm{dBm}$ \\
\hline
Obstacle altitude $H_{\text{obs}}$ & $120\ \mathrm{m}$ \\
\hline
UAV's maximum acceleration $a_{\max}$ & $5\ \mathrm{m/s^2}$ \\
\hline
UAV's maximum speed $v_{\max}$ & $20\ \mathrm{m/s}$ \\
\hline
Blade power $P_1$ & $59.03\ \mathrm{W}$ \\
\hline
Hovering power $P_2$ & $79.07\ \mathrm{W}$ \\
\hline
Rotor average speed $v_0$ & $3.6\ \mathrm{m/s}$ \\
\hline
Blade tip speed $U_{\text{tip}}$ & $120\ \mathrm{m/s}$ \\
\hline
Rotor area $A_0$ & $0.5030\ \mathrm{m^2}$ \\
\hline
Cylindrical obstacle's safety distance $d_{\text{safe}}^c$ & $10\ \mathrm{m}$ \\
\hline
UAV's safe distance $d_{\min}$ & $3\ \mathrm{m}$ \\
\hline
Rectangular obstacle's safety distance $\mathbf{d}_{\text{safe}}^r$ & $[10,10]\ \mathrm{m}$ \\
\hline
\end{tabular}
\end{table}
	\subsection{Simulation Setup}
In this section, we evaluate the performance of the proposed algorithm through a series of comprehensive simulations. The simulation environment is configured as a horizontal rectangular area of size $500~\mathrm{m} \times 500~\mathrm{m}$, within which both the UDs and UAVs are located. In the vertical dimension, the UDs are randomly distributed on the ground, while the UAVs are deployed at altitudes ranging from $[100,150]~\mathrm{m}$. 

    The DRL framework is implemented using PyTorch 2.3.1 with the Tianshou library, and it is trained on an NVIDIA RTX 4070 GPU under Windows with Python 3.11.10. The hyperparameters used in the implementation are summarized below: The EL is set to 500, and $e^{\text{MAX}}$ is 1000. The replay buffer size is $5 \times 10^4$, and the learning rate is $5 \times 10^{-4}$. The discount factor $\gamma$ is 0.99, and the batch size is 256. The hidden layers have sizes $[256, 128]$, and the optimizer is Adam. Each UAV is initialized with an onboard energy capacity of $20,000~\mathrm{J}$~\cite{10813002}, and their initial positions are randomly assigned within the designated area. The remaining environment parameters follow the configurations outlined in~\cite{10130436}, and are summarized in Table~\ref{tab:parameters}.

    \subsection{Performance Evaluation}
    	\begin{figure}[t]
		\centering
		\subfigure[Performance comparison of different algorithms.]{
			\includegraphics[width=0.45\textwidth]{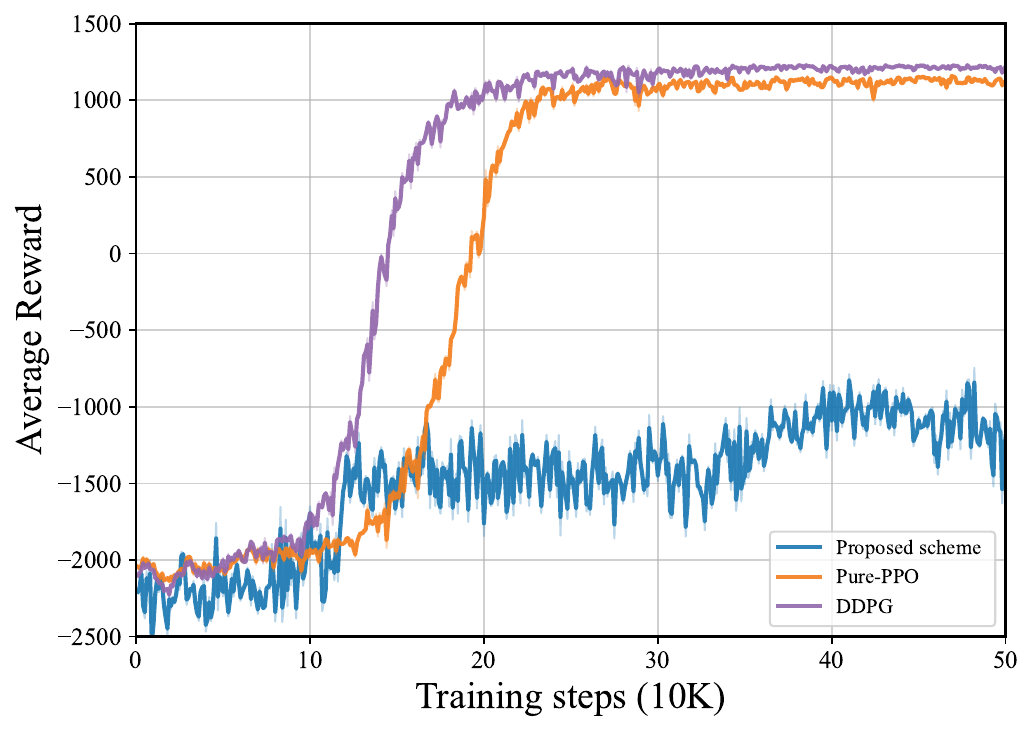}
		}
		\subfigure[Performance comparison of different random seeds.]{
			\includegraphics[width=0.45\textwidth]{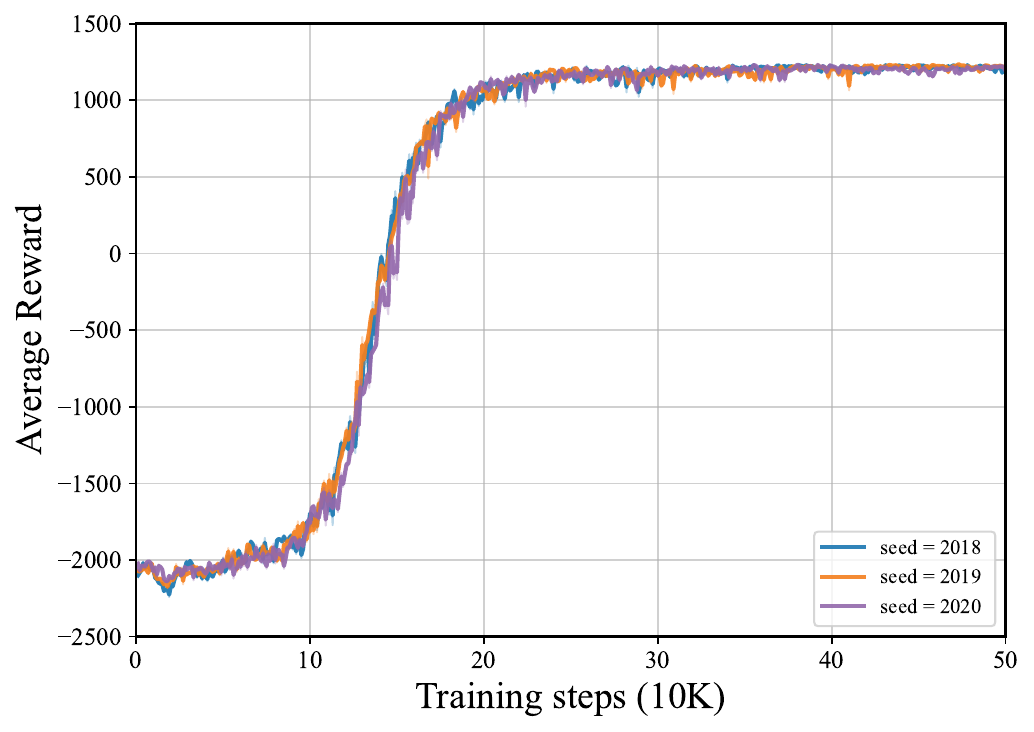}
		}
		\caption{The performance evaluation of the proposed scheme.}
		\label{fig:convergen}
	\end{figure}
    \textit{1) Convergence:} As illustrated in Fig.~\ref{fig:convergen}(a), we compare the proposed PPO-based scheme with the Pure-PPO and DDPG baselines in terms of user satisfaction throughout the training process. It is evident that the PPO framework achieves notably higher learning efficiency than DDPG. This advantage primarily stems from the surrogate objective function adopted by PPO, which facilitates more stable policy updates, as well as its stochastic policy structure, which promotes effective exploration. Furthermore, compared with the Pure-PPO baseline, the proposed PPO-based scheme integrates a water-filling algorithm for beamforming prior to the training stage. This pre-processing step significantly reduces the dimensionality of the action space, thereby improving computational efficiency and accelerating convergence. As a result, the proposed method converges effectively after approximately 150K training steps, while Pure-PPO requires around 200K steps to exhibit convergence. In contrast, DDPG fails to demonstrate stable convergence over the entire training period. Overall, the proposed PPO-based scheme outperforms the baseline algorithms in terms of both final performance and training stability.
    
   In general, the generalization capability serves as a critical criterion for evaluating the performance of DRL algorithms. Fig.~\ref{fig:convergen}(b) illustrates the convergence of the average reward under different training seeds. Once the learning process stabilizes, the reward values converge to a similar range across different seeds, indicating that the learned policy exhibits consistent performance under diverse initialization conditions.
    
    	\begin{figure}[t]
		\centering
		\includegraphics[width=0.9\columnwidth]{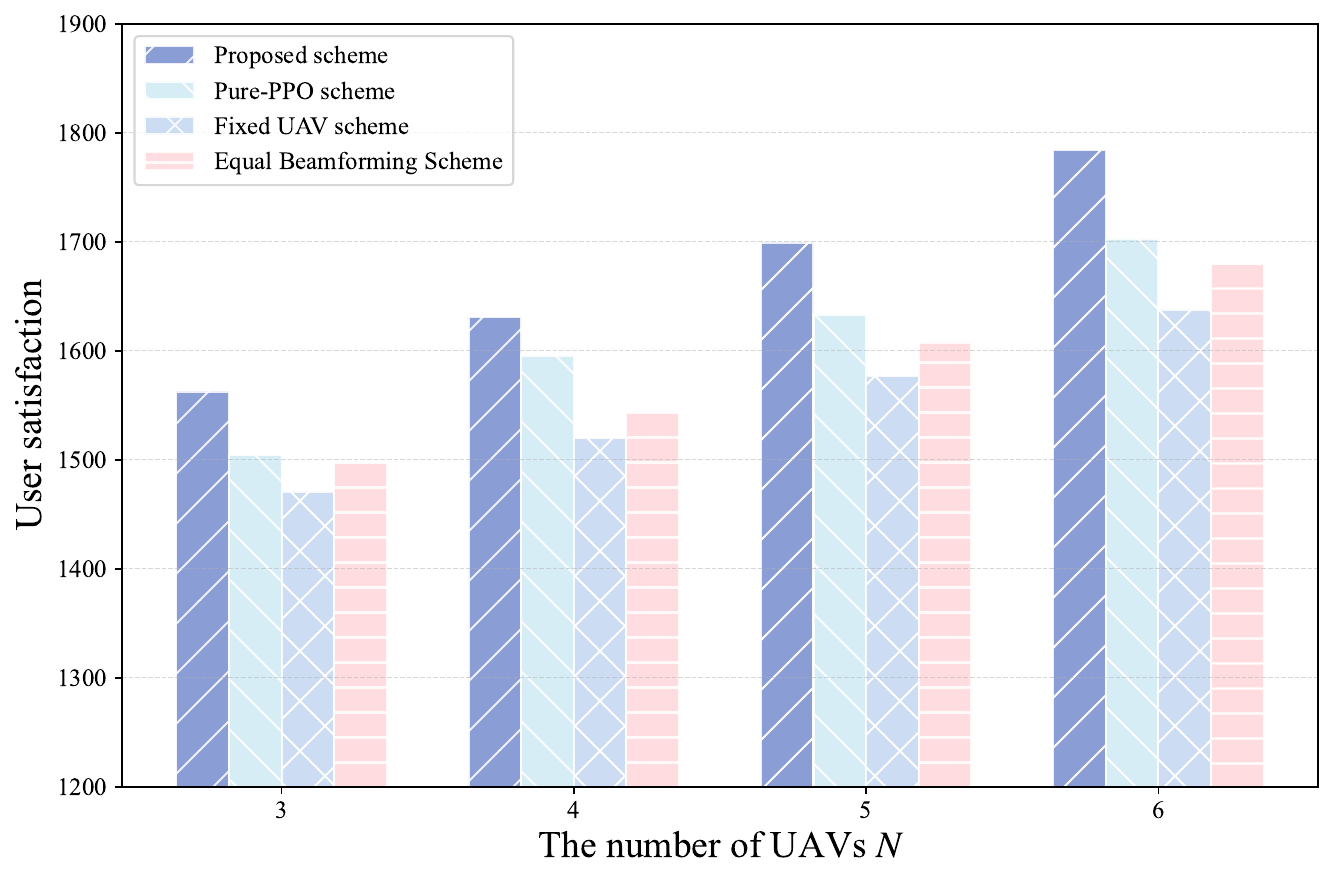}
		\caption{The user satisfaction versus the number of UAVs $N$.}
		\label{fig:uav_num}
	\end{figure}
    
        	\begin{figure}[t]
		\centering
		\includegraphics[width=0.9\columnwidth]{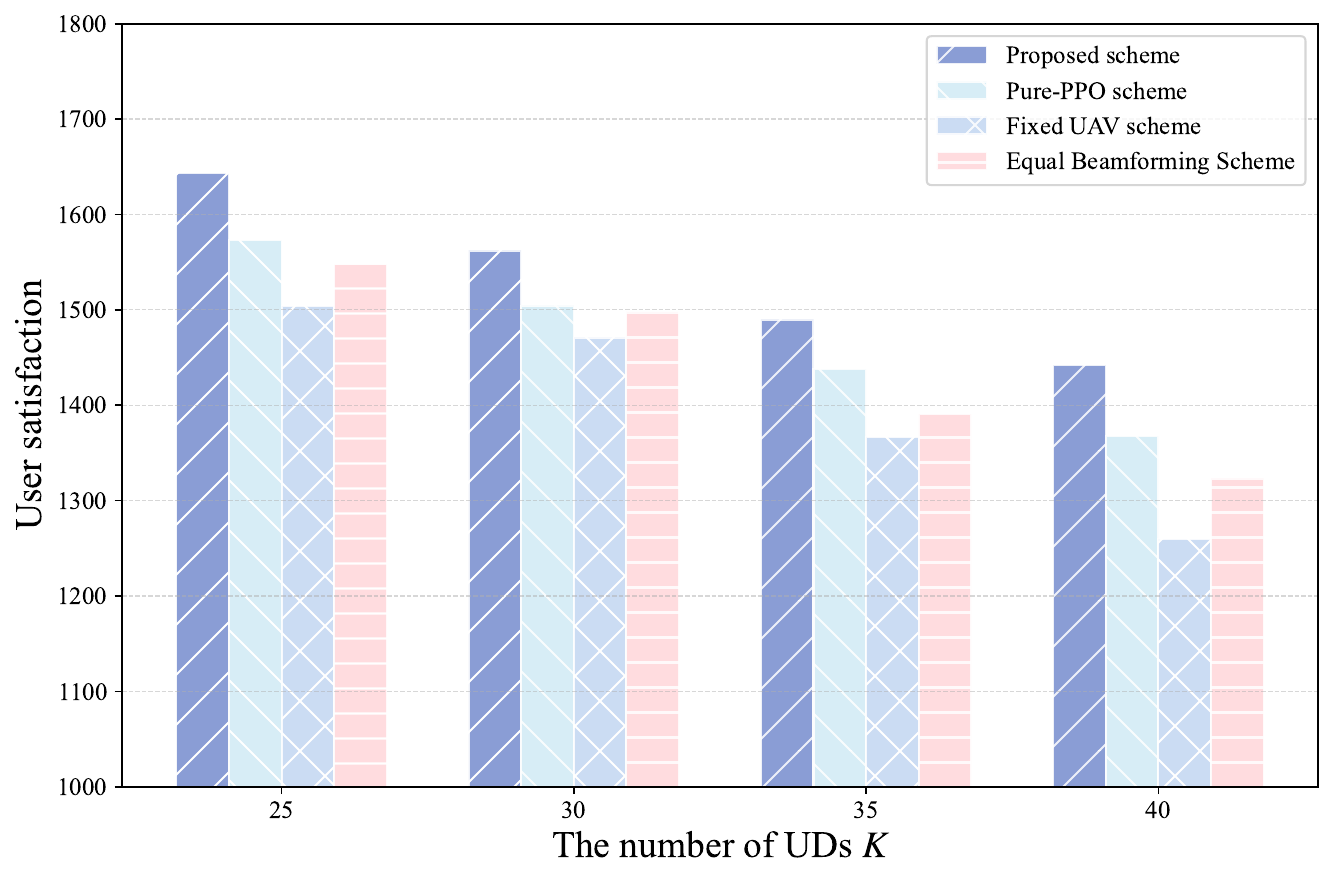}
		\caption{The user satisfaction versus the number of UDs $K$.}
		\label{fig:user_num}
	\end{figure}
    \textit{2) Scalability: }Fig.~\ref{fig:uav_num} presents the performance comparison of four baseline schemes under different numbers of UAVs with $K$ = 30. It can be observed that user satisfaction increases steadily as the number of UAVs grows. This improvement stems from the fact that a larger number of UAVs enables the system to deliver more personalized services to each UD, thereby enhancing communication quality and ultimately improving user satisfaction. Moreover, the proposed PPO-based scheme consistently outperforms all baseline methods across all evaluated configurations.

    In addition, we further compare the performance of various schemes under different numbers of UDs. To mitigate the impact of randomness in UD distribution, the random seed is fixed to 2018 across all schemes, ensuring consistency in UD placement across varying UD counts. As illustrated in Fig.~\ref{fig:user_num}, when the number of UDs is relatively small, all schemes achieve comparable performance in terms of user satisfaction, approaching the optimal level. This is mainly because, in such scenarios, the ZF technique can fully leverage the available spatial degrees of freedom to effectively eliminate inter-UD interference. It is also evident that the proposed PPO-based method maintains superior performance across all UD densities, further demonstrating its robustness and scalability.
    
	\begin{figure}[t]
		\centering
		\includegraphics[width=0.9\columnwidth]{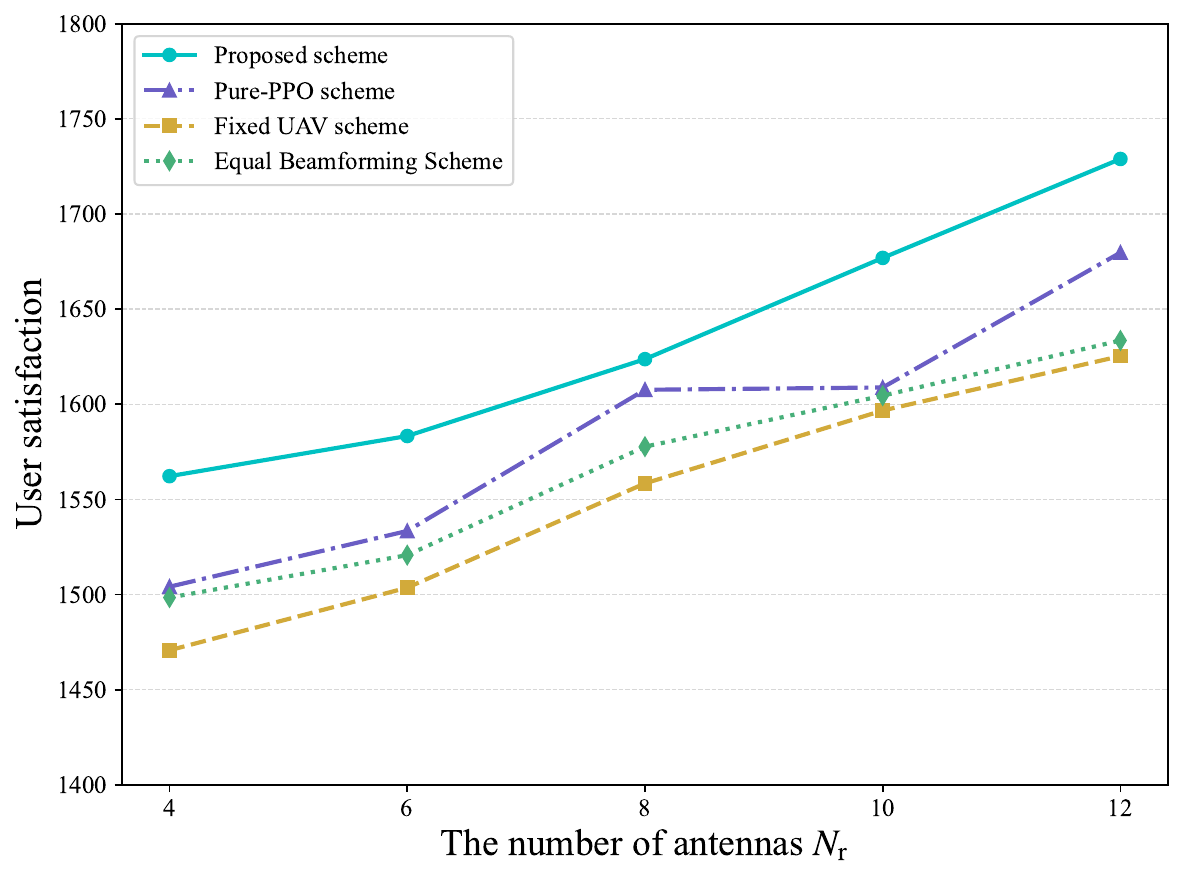}
		\caption{The user satisfaction versus the number of antennas $N_{\mathrm{r}}$.}
		\label{fig:antenna_num}
	\end{figure}
    \textit{3) Sensitivity:}
    Fig.~\ref{fig:antenna_num} illustrates the performance gain achieved by the proposed PPO-based scheme under varying numbers of UAV antennas. The system parameters are configured with $K$ = 30 UDs and $N$ = 3 UAVs. As depicted in the figure, user satisfaction improves steadily as the number of UAV antennas increases. This trend is primarily attributed to the fact that, under the ZF precoding strategy, a larger number of antennas provides enhanced spatial degrees of freedom and stronger interference suppression capabilities, thereby improving the effective channel gain. Additionally, in multi-UD scenarios, the availability of more antennas allows for more precise beam allocation to individual UDs, which further mitigates inter-UD interference and ultimately contributes to higher user satisfaction.
            \begin{figure}[t]
		\centering
		\includegraphics[width=0.9\columnwidth]{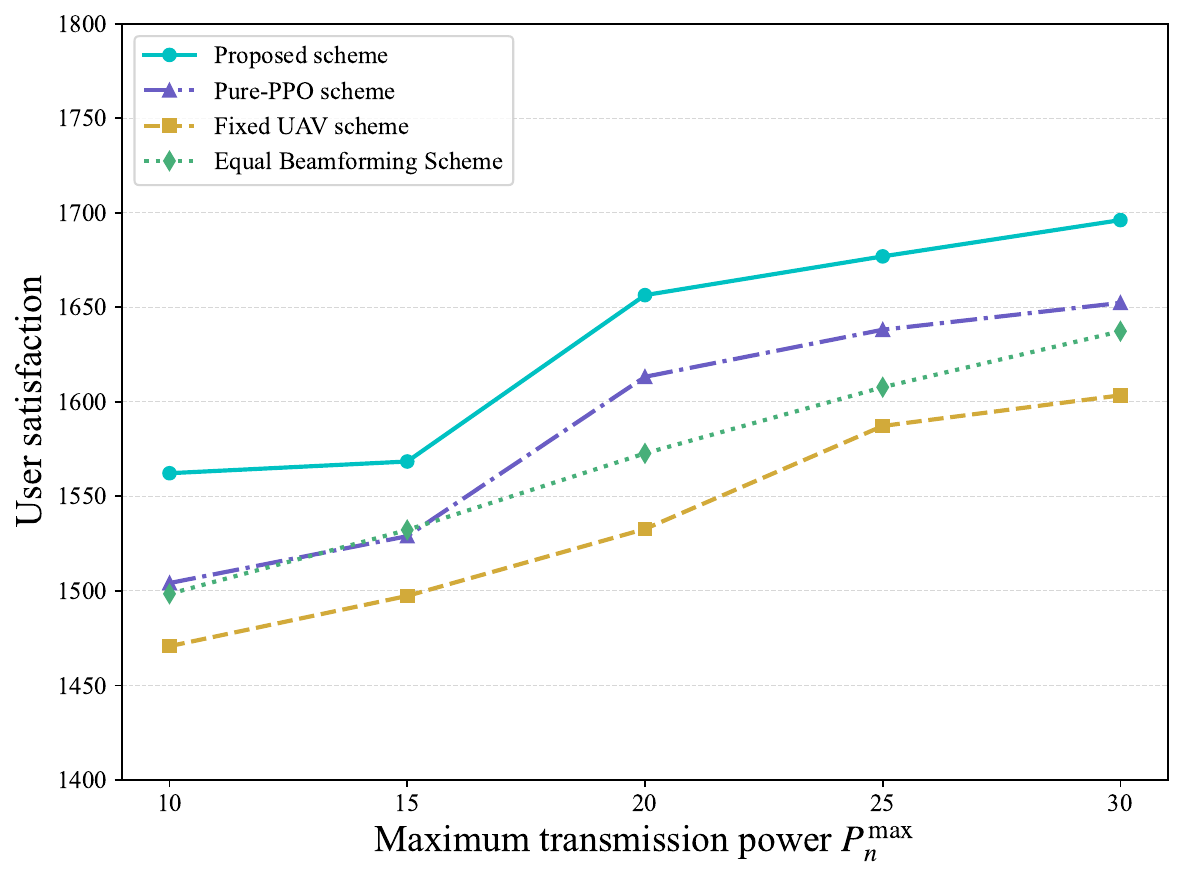}
		\caption{The user satisfaction versus the maximum power of UAV $P_n^{\max}$.}
		\label{fig:power_max}
	\end{figure}
    
     Fig.~\ref{fig:power_max} further compares the user satisfaction achieved by different schemes under identical antenna configurations but varying maximum transmit power levels. It is clearly observed that, as the transmit power increases, the performance gain of the proposed scheme becomes increasingly evident relative to the baseline methods. This result confirms the effectiveness of jointly optimizing UAV trajectories and UAV-UD association indicators in improving overall spectral efficiency. Notably, under sufficient transmit power conditions, the performance advantage of the proposed optimization strategy becomes significantly more pronounced.

    \begin{figure}[t]
		\centering
		\includegraphics[width=0.9\columnwidth]{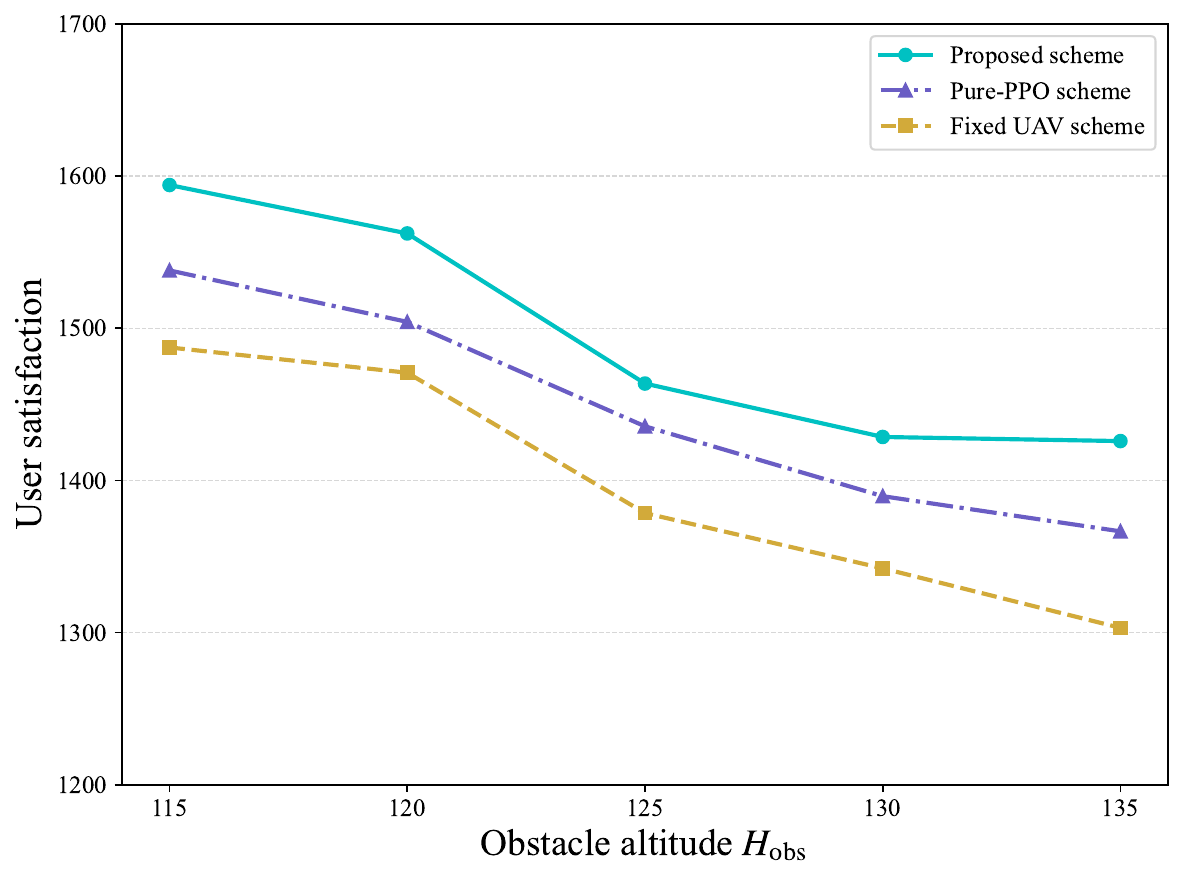}
		\caption{The user satisfaction versus the altitude of obstacle $H_\text{obs}$.}
		\label{fig:obs_altitude}
	\end{figure}

    Fig.~\ref{fig:obs_altitude} presents the performance comparison of various schemes under different obstacle height settings. It is observed that user satisfaction declines for all schemes as obstacle height increases. This degradation is primarily due to the heightened risk of UAV-obstacle collisions and the corresponding reduction in available maneuvering space, which collectively impair the UAVs' ability to deliver personalized services. Despite this adverse impact, the proposed PPO-based scheme consistently outperforms all baseline methods across the entire range of obstacle heights, demonstrating its robustness and adaptability in complex environments.
     \begin{figure}[t]
		\centering
		\includegraphics[width=0.9\columnwidth]{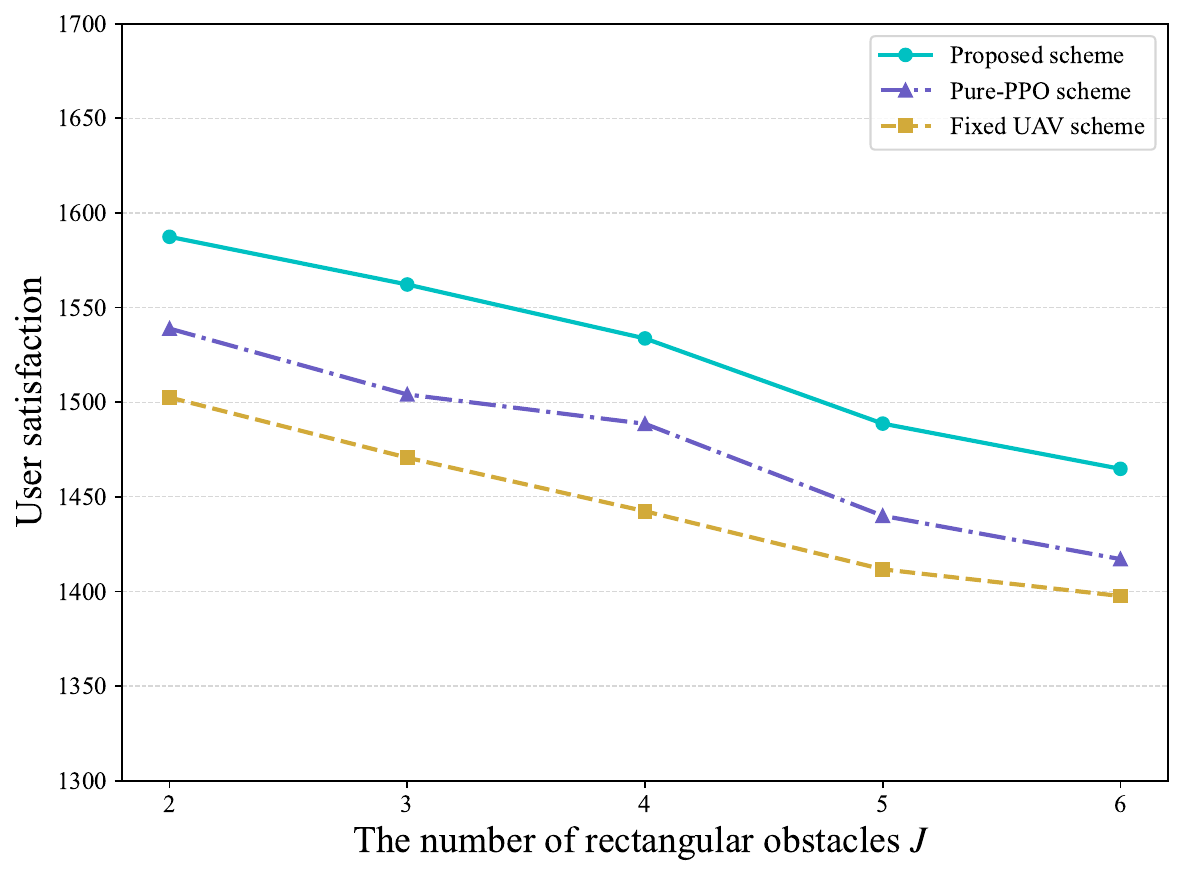}
		\caption{The user satisfaction versus the number of rectangular obstacles $J$.}
		\label{fig:obs_num}
	\end{figure}
    Similarly, Fig.~\ref{fig:obs_num} demonstrates that the overall user satisfaction of all schemes decreases with an increasing number of rectangular obstacles. A higher obstacle density not only imposes more stringent constraints on feasible UAV flight trajectories but also raises the likelihood of NLoS conditions, both of which degrade communication quality and reduce the system’s ability to meet heterogeneous user demands. As cylindrical obstacles are relatively uncommon in practical deployments and exhibit blocking effects comparable to those of rectangular obstacles, their separate evaluation is excluded from this study.

    	\begin{figure}[t]
		\centering
		\subfigure[UAV trajectory under the proposed PPO scheme.]{
			\includegraphics[width=0.2\textwidth]{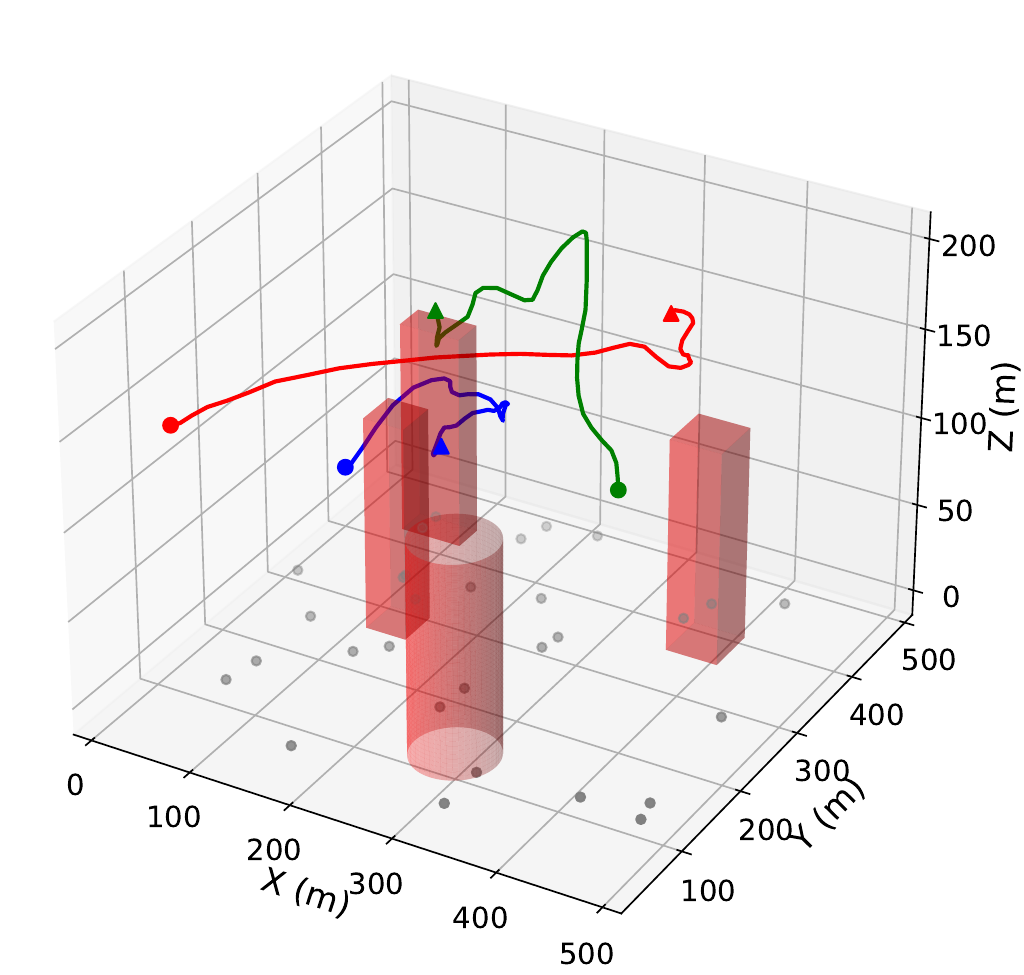}
		}
    \includegraphics[width=0.049\textwidth]{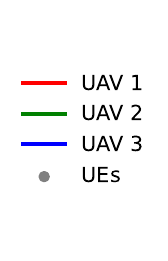}
		\subfigure[UAV trajectory under Pure-PPO scheme.]{
			\includegraphics[width=0.2\textwidth]{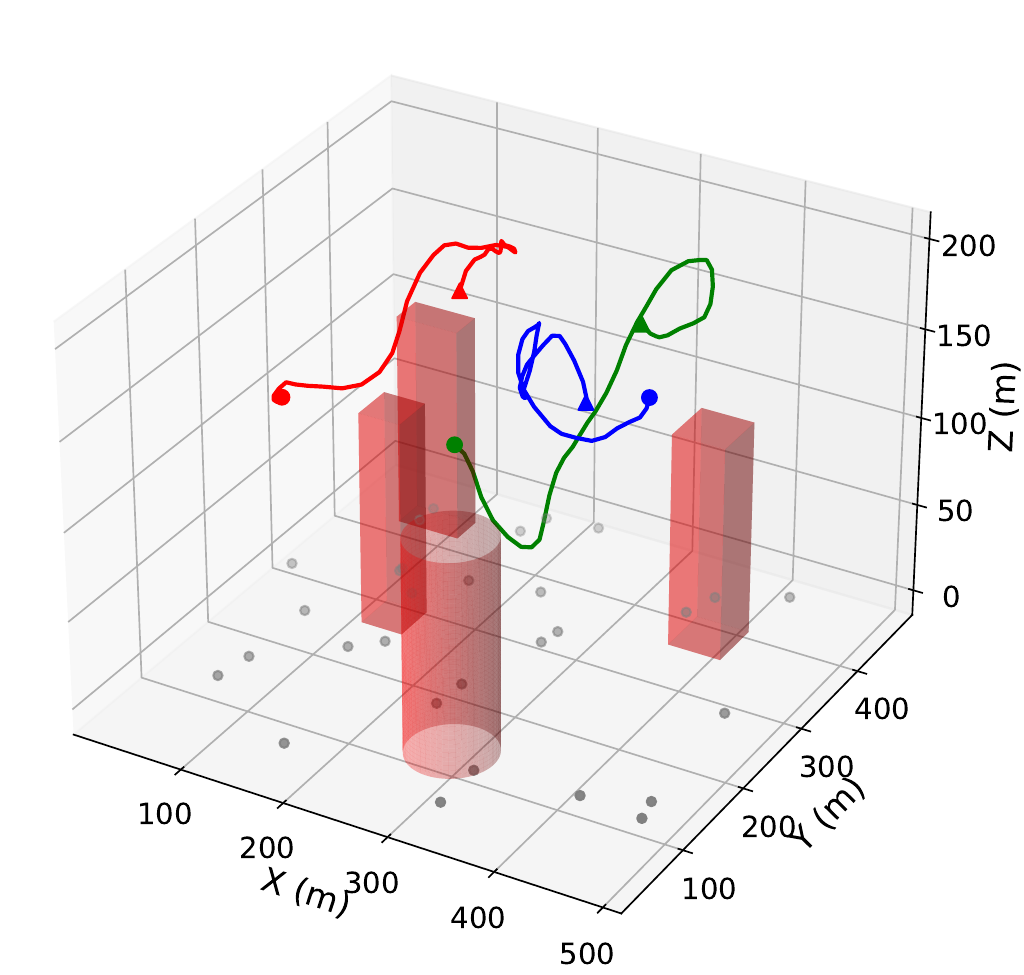}
		}
		\caption{Example of 3D trajectory of the UAV, where $N=3$ and $K=30$.}
		\label{fig:trajectories}
	\end{figure}
       
   \textit{4) Trajectories:} Fig.~\ref{fig:trajectories} illustrates the 3D trajectories of UAVs under two different schemes. As shown in the figure, the UAVs tend to navigate toward areas with high UD density while effectively avoiding obstacles. In scenarios where certain UDs are located within or near obstacle regions, the UAVs approach these areas to maintain uninterrupted communication services. Notably, the UAVs attempt to minimize their distance to the UDs, as a shorter UAV-UD link results in lower path loss. Additionally, a moderate increase in flight altitude enables the UAVs to circumvent obstacles and simultaneously improve the elevation angle to the UDs, thereby enhancing channel conditions and achieving higher data rates. In Fig.~\ref{fig:trajectories}(a), the UAVs demonstrate more flexible and dynamic trajectory planning, actively exploring wider regions to optimize coverage. In contrast, as shown in Fig.~\ref{fig:trajectories}(b), the Pure-PPO scheme exhibits more conservative behavior, characterized by prolonged hovering in limited areas and less adaptive exploration.
   
	\section{Conclusion}
This paper has presented a novel UAV-assisted communication framework designed to provide efficient and reliable services in heterogeneous networks with complex ground obstacles. By jointly optimizing the UAV 3D trajectories, beamforming vectors, and UAV-UD association indicators, the objective has been to maximize overall user satisfaction. To effectively handle the complexity of the problem, a BCD approach has been adopted to decouple it into two subproblems, which have subsequently been solved using a water-filling algorithm and the PPO algorithm, respectively. Simulation results have demonstrated that the proposed scheme significantly enhances user satisfaction, accelerates convergence, and achieves effective obstacle avoidance. Future work will investigate the impact of dynamic obstacles on trajectory planning and explore the robustness of the proposed scheme in high-dynamic environments.

\bibliographystyle{IEEEtran}
\bibliography{reference.bib}

\end{document}